\documentclass[journal,11pt,onecolumn]{IEEEtran}

\usepackage[noadjust]{cite}

\ifCLASSINFOpdf
\else
\usepackage[dvips]{graphicx,color}
\fi
\usepackage[cmex10]{amsmath}
\usepackage{amssymb}
\newtheorem{definition}{Definition}[section]
\newtheorem{theorem}{Theorem}[section]
\newtheorem{lemma}{Lemma}[section]
\newtheorem{remark}{Remark}[section]

\newtheorem{corollary}{Corollary}[section]
\usepackage[T1]{fontenc}
\usepackage{ae}
\usepackage{aecompl}
\makeatletter
\renewcommand{\theequation}{\arabic{section}.\arabic{equation}}
\@addtoreset{equation}{section}
\makeatother
\makeatletter
    \renewcommand{\thefigure}{%
    \arabic{section}.\arabic{figure}}
\@addtoreset{figure}{section}
\makeatother

\begin{document}
%
\title{First- and Second-Order Hypothesis Testing \\ for Mixed Memoryless Sources with \\ General Mixture}
%
%
%
\author{
        Te~Sun~Han,~\IEEEmembership{Life~Fellow,~IEEE,}
        and Ryo Nomura,~\IEEEmembership{Member,~IEEE}
\thanks{T. S. Han is with the National Institute of Information and Communications Technology (NICT), Tokyo, Japan, 
e-mail: han@is.uec.ac.jp }
\thanks{R. Nomura is with the School of Network and Information, Senshu University, Kanagawa, Japan,
 e-mail: nomu@isc.senshu-u.ac.jp}
}

\maketitle
\begin{abstract}
The {\it first- and second-order} optimum achievable exponents in the simple hypothesis testing problem are investigated. The optimum achievable exponent for type II error probability, under the constraint that the type I error probability is allowed asymptotically up to $\varepsilon$, is called the $\varepsilon$-optimum exponent. 
In this paper, we first give the \textit{second-order} $\varepsilon$-exponent in the case where the \textit{null} hypothesis and the \textit{alternative} hypothesis are a mixed memoryless source  and a stationary memoryless source, respectively.
We next generalize this setting to the case where the \textit{alternative} hypothesis is also a mixed memoryless source. 
We address the \textit{first-order} $\varepsilon$-optimum exponent in this setting.
In addition, an extension of our results to more general setting such as the hypothesis testing with mixed \textit{general} source and the relationship with the \textit{general} compound hypothesis testing problem are also discussed.
\end{abstract}

\begin{IEEEkeywords}
General Source, Hypothesis Testing, Information Spectrum, Mixed Source, Optimum Exponent
\end{IEEEkeywords}

%
\IEEEpeerreviewmaketitle

%
%
%
%


\section{Introduction}
Let ${\bf X} = \{X^n \}_{n=1}^\infty$ 
and $\overline{{\bf X}} = \{\overline{X}^n \}_{n=1}^\infty$ be two \textit{general sources}, where  
we use the term of \textit{general source} to denote a sequence of random variables $X^n$ (resp. $\overline{X}^n$) indexed by block length $n$, where each component of $X^n$  (resp. $\overline{X}^n$) may vary depending on $n$.

We consider the hypothesis testing problem with null hypothesis ${\bf X}$, alternative hypothesis $\overline{\bf X}$ and acceptance region ${\cal A}_n \subset {\cal X}^n$. 
The probabilities of type I error and type II error are defined, respectively, as
\begin{eqnarray*}
\mu_n := \Pr\left\{ X^n \notin {\cal A}_n\right\}, \quad \lambda_n := \Pr\left\{ \overline{X}^n \in {\cal A}_n\right\}.
\end{eqnarray*}
In this paper, we focus mainly on how to determine the $\varepsilon$-optimum exponent, defined as  the supremum of achievable exponents $R$ for the type II error probability $\lambda_n \simeq e^{-nR}$ under the constraint that the type I error probability is allowed asymptotically up to a constant $\varepsilon$ $(0 \le \varepsilon <1)$.
The fundamental result in this setting is so-called Stein's lemma \cite{DZ}, which gives the $\varepsilon$-optimum exponent in the case where the \textit{null} and \textit{alternative} hypotheses are stationary memoryless sources. 
The lemma shows that the $\varepsilon$-optimum exponent is given by $D(P_X||P_{\overline{X}})$, the divergence between stationary memoryless sources $X$ and $\overline{X}$.
Chen \cite{CA96} has generalized this lemma to the case where both of ${\bf X}$ and $\overline{\bf X}$ are \textit{general} sources, and established the \textit{general} formula of $\varepsilon$-optimum exponent in terms of \textit{divergence spectra}.
The $\varepsilon$-optimum exponent derived by him is called in this paper the \textit{first-order} $\varepsilon$-optimum exponent.

On the other hand, achievable rates called those of the {\it second-order} have been investigated in several contexts in information theory \cite{Strassen,Hayashi2,Polyanskiy2010,NH2011,TK2014,W2017} in order to investigate \textit{finer} asymptotic behaviors of the information-theoretic quantities.
Strassen \cite{Strassen} has first introduced the notion of the $\varepsilon$-optimum achievable exponent of the \textit{second-order} in the hypothesis testing problem, called the \textit{second-order} $\varepsilon$-optimum exponent, and derived the \textit{second-order} $\varepsilon$-optimum exponent in the case where ${\bf X}$ and $\overline{\bf X}$ are stationary memoryless sources.
Han \cite{Han_Jeju} has demonstrated the \textit{general} formula (though not single-letterized)  of  the \textit{second-order} $\varepsilon$-optimum exponent. 
The results in \cite{Strassen} and \cite{Han_Jeju} also have revealed that the asymptotic normality of the \textit{divergence density rate} (or the \textit{likelihood ratio rate}) plays an important role also in computing the \textit{second-order} $\varepsilon$-optimum exponent.

In this paper, on the other hand, we consider the hypothesis testing for \textit{mixed} memoryless sources in which the asymptotic normality of \textit{divergence density rate} does not hold.
The class of mixed sources is quite important, because all of stationary sources can be regarded as forming mixed sources consisting of stationary ergodic sources.
Therefore, the analysis for mixed sources is insightful and so we first focus on the case with \textit{mixed} memoryless source ${\bf X}$.
In this direction, Han \cite{Han} has first derived the formula for the \textit{first-order} $\varepsilon$-optimum exponent in the case with mixed memoryless source ${\bf X}$ and stationary memoryless source $\overline{\bf X}$.
One of our first main results is to establish the \textit{second-order} $\varepsilon$-optimum exponent in the same setting by invoking the relevant asymptotic normality. 
The result is a substantial generalization of that of Strassen \cite{Strassen}.
Second, we generalize this setting to the case where both of \textit{null} and \textit{alternative} hypotheses are mixed memoryless ${\bf X}$, $\overline{\bf X}$ to establish the \textit{first-order} $\varepsilon$-optimum exponent.

It should be emphasized that our results described here are valid for mixed memoryless sources with \textit{general mixture} in the sense that the mixing weights may be arbitrary probability measures.
For the case of mixed memoryless sources (or also mixed \textit{geneal} sources) with \textit{discrete mixture}, we can reveal the deep relationship with the \textit{compound} hypothesis testing problem.
We notice that the \textit{compound} hypothesis testing problem is important from both theoretical and practical points of view.
We show that \textit{first-order} $0$-optimum (resp. exponentially $r$-optimum) exponents for the mixed general hypothesis testing coincide with those for the $0$-optimum (resp. exponentially $r$-optimum) exponents in the \textit{compound} general hypothesis testing.

The present paper is organized as follows. In Section II,
we define the problem setting and review the \textit{general formulas} for the \textit{first}- and \textit{second-order} $\varepsilon$-optimum exponents. 
In Section III, we establish the \textit{second-order} $\varepsilon$-optimum exponents in the case with mixed memoryless source ${\bf X}$ and stationary memoryless source $\overline{\bf X}$.
In Section IV, we consider the case where both of \textit{null} and \textit{alternative} hypotheses are mixed memoryless sources, and derive the \textit{first}-order $\varepsilon$-optimum exponent.
Section V is devoted to an extension of mixed memoryless sources to mixed \textit{general} sources.
Finally, in Section VI we define the optimum exponent for the \textit{compound} general hypothesis testing problem and discuss the relevant relationship with the hypothesis testing with \textit{mixed} general sources.
We conclude the paper in Section VII.
\section{General formulas for $\varepsilon$-hypothesis testing}
We first review the first-order \textit{general formula} and derive the second-order \textit{general formula}.
Throughout in this paper, the following lemmas play the important role and 
 we use the notation that $P_Z$ indicates the probability distribution of random variable $Z$.
\begin{lemma}[{\cite[Lemma 4.1.1]{Han}}] \label{upper}
For any $t>0$, define the acceptance region as 
\[
{\cal A}_n = \left\{ {\bf x} \in {\cal X}^n \left| \frac{1}{n}\log\frac{P_{{X}^n}({\bf x})}{P_{\overline{X}^n}({\bf x})} \ge t \right. \right\},
\]
then, it holds that
\[
\Pr \left\{ \overline{X}^n \in {\cal A}_n \right\} \le e^{-nt}.
\]
\end{lemma}
\begin{lemma}[{\cite[Lemma 4.1.2]{Han}}] \label{lower}
For any $t>0$ and any ${\cal A}_n$, it holds that
\[
\Pr \left\{ {X}^n \notin {\cal A}_n \right\} + e^{nt}\Pr \left\{ \overline{X}^n \in {\cal A}_n \right\}  \ge \Pr \left\{ \frac{1}{n} \log \frac{P_{{X}^n}({X^n})}{P_{\overline{X}^n}({X^n})} \le t \right\}.
\]
\end{lemma}
Although the proof of these lemmas is simple and found in \cite{Han}, we record it in Appendix A for self-containedness.

We define the \textit{first} and \textit{second}-order $\varepsilon$-optimum exponents as follows.
\begin{definition}
Rate $R$ is said to be $\varepsilon$-achievable, if there exists an acceptance region  ${\cal A}_n$ such that
\begin{equation} \label{eq:1st_ach}
\limsup_{n \to \infty} \mu_n \le \varepsilon \   \mbox{ and } \ \liminf_{n \to \infty} \frac{1}{n}\log \frac{1}{\lambda_n} \ge R.
\end{equation}
\end{definition}
\begin{definition}[First-order $\varepsilon$-optimum exponent]
\begin{equation}
B_{\varepsilon}({\bf X}||\overline{\bf X}) := \sup\{ R | R \mbox{ is $\varepsilon$-achievable} \}.
\end{equation}
\end{definition}
The right-hand side of (\ref{eq:1st_ach}) specifies the asymptotic behavior of the form $\lambda_n \simeq e^{-nR}$.
Chen \cite{CA96} has derived the general formula for $B_{\varepsilon}({\bf X}||\overline{\bf X})$:
\begin{theorem}[Chen \cite{CA96}]  \label{theo:general_1st}
\begin{equation}
B_{\varepsilon}({\bf X}||\overline{\bf X}) = \sup\{ R | K(R) \le \varepsilon \} \quad (0 \le \forall\varepsilon <1),
\end{equation}
where 
\[
K(R) = \limsup_{n \to \infty} \Pr\left\{ \frac{1}{n}\log \frac{P_{X^n}(X^n)}{P_{\overline{X}^n}(X^n)} \le R\right\}.
\]
\end{theorem}
\begin{IEEEproof} 

The proof is found in \cite{Han} and similar to that of Theorem \ref{theo:general_2nd} below.
\end{IEEEproof}

\begin{definition}
Rate $S$ is said to be $(\varepsilon,R)$-achievable, if there exists an acceptance region  ${\cal A}_n$ such that
\begin{equation}  \label{eq:2nd_ach}
\limsup_{n \to \infty} \mu_n \le \varepsilon \   \mbox{ and } \ \liminf_{n \to \infty} \frac{1}{\sqrt{n}}\log \frac{1}{\lambda_n e^{nR}} \ge S.
\end{equation}
\end{definition}
\begin{definition}[Second-order $(\varepsilon,R)$-optimum exponent]
\begin{equation}
B_{\varepsilon}(R|{\bf X}||\overline{\bf X}) := \sup\{ S | S \mbox{ is $(\varepsilon,R)$-achievable} \}.
\end{equation}
\end{definition}
The right-hand side of (\ref{eq:2nd_ach}) specifies the asymptotic behavior of the form $\lambda_n \simeq e^{-nR - \sqrt{n}S}$.
Han \cite{Han_Jeju} has derived the general formula for $B_{\varepsilon}(R|{\bf X}||\overline{\bf X})$:
\begin{theorem}[Han \cite{Han_Jeju}] \label{theo:general_2nd}
\begin{equation}
B_{\varepsilon}(R|{\bf X}||\overline{\bf X}) = \sup\{ S | K(R,S) \le \varepsilon \} \quad (0 \le \forall\varepsilon <1),
\end{equation}
where 
\[
K(R,S) = \limsup_{n \to \infty} \Pr\left\{ \frac{1}{n}\log \frac{P_{X^n}(X^n)}{P_{\overline{X}^n}(X^n)} \le R + \frac{S}{\sqrt{n}}\right\}.
\]
\end{theorem}
\begin{IEEEproof}

The proof consists of two parts.

1) \textit{Direct Part:}

Set $S_0 =  \sup\{ S | K(R,S) \le \varepsilon \}$. Then, we show that $S = S_0 -\gamma$ is $(\varepsilon,R)$-achievable for $\forall \gamma>0$.

Define the acceptance region ${\cal A}_n$ as 
\[
{\cal A}_n = \left\{ \frac{1}{n} \log \frac{P_{X^n}({\bf x})}{P_{\overline{X}^n}({\bf x})}  > R + \frac{S}{\sqrt{n}}\right\}.
\]
Then, from Lemma \ref{upper} with $t = R + \frac{S}{\sqrt{n}}$ we have the upper bound for the type II error probability $\lambda_n$:
\[
\lambda_n = \Pr \left\{ \overline{X}^n \in {\cal A}_n \right\} \le e^{-nR - \sqrt{n}S},
\]
from which it follows that
\begin{equation} \label{eq:1-1}
\liminf_{n \to \infty} \frac{1}{\sqrt{n}} \log \frac{1}{\lambda_n e^{nR}} \ge S.
\end{equation}
We next evaluate the type I error probability $\mu_n$. Noting that 
\[
\mu_n =  \Pr\left\{ X^n \notin {\cal A}_n \right\} =
\Pr\left\{ \frac{1}{n}\log \frac{P_{X^n}(X^n)}{P_{\overline{X}^n}(X^n)} \le R + \frac{S}{\sqrt{n}}\right\},
\]
we have
\begin{equation} \label{eq:1-2}
\limsup_{n \to \infty} \mu_n = \limsup_{n \to \infty} \Pr\left\{ \frac{1}{n}\log \frac{P_{X^n}(X^n)}{P_{\overline{X}^n}(X^n)} \le R + \frac{S}{\sqrt{n}}\right\} \le \varepsilon,
\end{equation}
because $S= S_0 -\gamma$ by the definition.
Hence, from (\ref{eq:1-1}) and (\ref{eq:1-2}), $S = S_0 -\gamma$ is $(\varepsilon,R)$-achievable.
Since $\gamma >0$ is arbitrary, the direct part has been proved.

2) \textit{Converse Part:}

Suppose that $S$ is $(\varepsilon,R)$-achievable.
Then, there exists an acceptance region ${\cal A}_n$ such that 
\begin{eqnarray}  \label{eq:2-0}
\limsup_{n \to \infty}\mu_n \le \varepsilon \ \mbox{ and } \ \liminf_{n \to \infty} \frac{1}{\sqrt{n}} \log \frac{1}{\lambda_n e^{nR}} \ge S.
\end{eqnarray}
We fix this acceptance region ${\cal A}_n$.
The second inequality means that for any $\gamma >0$
\begin{equation} \label{eq:2-1}
\lambda_n \le e^{-nR - \sqrt{n}(S-\gamma)}
\end{equation}
holds for sufficiently large $n$.
On the other hand, from Lemma \ref{lower} with $t = R+ \frac{S - 2\gamma}{\sqrt{n}}$ it holds that
\[
\mu_n + e^{nR +\sqrt{n}(S-2\gamma)} \lambda_n \ge \Pr \left\{ \frac{1}{n} \log \frac{P_{{X}^n}(X^n)}{P_{\overline{X}^n}(X^n)} \le R+ \frac{S - 2\gamma}{\sqrt{n}} \right\}.
\]
Substituting (\ref{eq:2-1}) into this inequality, we have
\[
\mu_n + e^{-\sqrt{n}\gamma} \ge \Pr \left\{ \frac{1}{n} \log \frac{P_{{X}^n}(X^n)}{P_{\overline{X}^n}(X^n)} \le R+ \frac{S - 2\gamma}{\sqrt{n}} \right\},
\]
for sufficiently large $n$.
Thus, we have
\[
\limsup_{n \to \infty }\mu_n \ge \limsup_{n \to \infty }\Pr \left\{ \frac{1}{n} \log \frac{P_{{X}^n}(X^n)}{P_{\overline{X}^n}(X^n)} \le R+ \frac{S - 2\gamma}{\sqrt{n}} \right\}.
\]
Here, from (\ref{eq:2-0}) we have
\[
\varepsilon \ge \limsup_{n \to \infty}\mu_n \ge \limsup_{n \to \infty }\Pr \left\{ \frac{1}{n} \log \frac{P_{{X}^n}(X^n)}{P_{\overline{X}^n}(X^n)} \le R+ \frac{S - 2\gamma}{\sqrt{n}} \right\},
\]
which means that
\[
S - 2\gamma \le B_{\varepsilon}(R|{\bf X}||\overline{\bf X}).
\]
Since $\gamma >0$ is arbitrarily, the proof of the converse part has been completed.
\end{IEEEproof}
%
%
%
%
%
%
%
%
\section{Mixed memoryless sources}
\subsection{First-order $\varepsilon$-optimum exponent}
In the previous section, we have reviewed the \textit{formula} for general hypothesis testings.
In this and subsequent sections, we consider special but insightful cases and compute the optimum exponents in the single-letterized form.
Let $\Theta$ be an arbitrary probability space with \textit{general} probability measure $w(\theta)\ (\theta \in \Theta)$.
Then, the hypothesis testing problem to be considered in this section is stated as follows.
\begin{itemize}
\item The null hypothesis is a mixed stationary memoryless source ${\bf X}= \{X^n \}_{n=1}^\infty$, that is, for $\forall {\bf x}=(x_1,\cdots,x_n) \in  {\cal X}^n$ 
\begin{equation} \label{eq:mix1}
P_{X^n}({\bf x}) = \int_\Theta P_{X^n_\theta}({\bf x}) dw(\theta),
\end{equation}
where $X_\theta^n$ is a stationary memoryless source for each $\theta \in \Theta$ and 
\[
P_{X^n_\theta}({\bf x}) = \prod_{i=1}^n P_{X_\theta}(x_i)
\]
with generic random variable $X_\theta \ (\theta \in \Theta)$ taking values in ${\cal X}$.
\item The alternative hypothesis is a stationary memoryless source $\overline{\bf X} = \left\{ \overline{X}^n\right\}_{n=1}^\infty$ with generic random variable $\overline{X}$ taking values in ${\cal X}$, that is,
\[
P_{\overline{X}^n}({\bf x}) = \prod_{i=1}^n P_{\overline{X}}(x_i).
\]
\end{itemize}
We assume ${\cal X}$ to be a \textit{finite} alphabet hereafter.
In order to treat this special case, first we introduce an expurgated parameter set on the basis of \textit{types}, where the type $T$ of sequence ${\bf x} \in {\cal X}^n$ is the empirical distribution of ${\bf x}$, that is, $T = (N(x|{\bf x})/n)_{x \in {\cal X}}$ with the number $N(x|{\bf x})$ of $i$ such that $x_i = x \ (i=1,2,\cdots,n)$.

Let $T_1, T_2, \cdots, T_{N_n}$ denote all possible types of sequences of length $n$. Then, it is well-known that
\begin{equation} \label{type}
N_n \le (n+1)^{|{\cal X}|}.
\end{equation}
Now for each ${\bf x}\in {\cal X}^n$, we define the set 
\begin{equation}
{\Theta}({\bf x}) := \left\{ \theta \in \Theta \left|   P_{X^n_\theta}({\bf x}) \le e^{\sqrt[4]{n}}P_{X^n}({\bf x}) \right. \right\}.
\end{equation}
Since $P_{X^n_\theta}$ is an i.i.d. source for each $\theta \in \Theta$, the set ${\Theta}({\bf x})$ depends only on the type $T_k$ of sequence ${\bf x}$, and therefore, we may write $\Theta(T_k)$ instead of $\Theta({\bf x})$.
Moreover, we define the set
\[
\Theta^\ast_n := \bigcap_{k=1}^{N_n} \Theta(T_k).
\]
Then, we have the following lemma:
\begin{lemma} \label{lemma:expurgate}
Let ${\bf X} = \{ X^n \}_{n=1}^\infty$ denote a mixed memoryless source defined in (\ref{eq:mix1}), then we have
\begin{equation}
\int_{\Theta_n^\ast} dw(\theta) \ge 1 - (n+1)^{|{\cal X}|}e^{-\sqrt[4]{n}}.
\end{equation}
\end{lemma}
\begin{IEEEproof}

Since $P_{X^n}({\bf x})$ is the expectation of $P_{X^n_\theta}({\bf x})$ with respect to $w(\theta)$, Markov's inequality guarantees that
\[
\Pr\left\{ \theta \in \Theta(T_k) \right\} \ge 1 - e^{-\sqrt[4]{n}} \quad (k=1,2,\cdots,N_n),
\]
from which, together with (\ref{type}), it follows that
\[
\Pr\left\{ \theta \in \Theta_n^\ast \right\} \ge 1 - (n+1)^{|{\cal X}|}e^{-\sqrt[4]{n}}.
\]
\end{IEEEproof}
Next, we introduce two lemmas.
\begin{lemma}[Upper Decomposition Lemma]  \label{lemma:upper_dec}
Let ${\bf X} = \{ X^n \}_{n=1}^\infty$ be a mixed memoryless source and $\overline{X} = \left\{ \overline{X}^n \right\}_{n=1}^\infty$ be an \textit{arbitrary} {general} source. Then, for any $\theta \in \Theta_n^\ast$ and any real $z_n$ it holds that
\begin{IEEEeqnarray}{rCl}
\Pr\left\{  \frac{1}{n} \log \frac{P_{X^n}(X_\theta^n)}{P_{\overline{X}^n}(X_\theta^n)} \le z_n \right\} {\le} \Pr\left\{  \frac{1}{n} \log \frac{P_{X_\theta^n}(X_\theta^n)}{P_{\overline{X}^n}(X_\theta^n)} \le z_n +  \frac{1}{{\sqrt[4]{n}}^3} \right\}.
\end{IEEEeqnarray}
\end{lemma}
\begin{IEEEproof}

Since $P_{X^n_\theta}({\bf x}) \le e^{\sqrt[4]{n}}P_{X^n}({\bf x})$ holds for $\forall \theta \in \Theta_n^\ast$, we have 
\begin{IEEEeqnarray*}{rCl}
\Pr\left\{   \frac{1}{n} \log  {P_{X^n}(X^n_\theta)} \le z_n \right\}  
& \le & \Pr\left\{  \frac{1}{n} \log  {P_{X^n_\theta}(X^n_\theta)}  - \frac{1}{\sqrt[4]{n}^3}\le z_n \right\} \\
& = & \Pr\left\{  \frac{1}{n} \log P_{X^n_\theta}(X_\theta^n) \le z_n +  \frac{1}{\sqrt[4]{n}^3} \right\} 
\end{IEEEeqnarray*}
for any $z_n$. By using this inequality with $z_n + \frac{1}{n}\log P_{\overline{X}^n}(X_\theta^n)$ instead of $z_n$, we have
\begin{IEEEeqnarray*}{rCl}
\Pr\left\{   \frac{1}{n} \log   \frac{P_{{X}^n}(X_\theta^n)}{P_{\overline{X}^n}(X_\theta^n)}  \le z_n \right\}  
& \le & \Pr\left\{  \frac{1}{n} \log  \frac{P_{{X}^n}(X_\theta^n)}{P_{\overline{X}^n}(X_\theta^n)}  \le z_n +  \frac{1}{\sqrt[4]{n}^3} \right\} 
\end{IEEEeqnarray*}
which completes the proof. 
\end{IEEEproof}

\begin{lemma}[Lower Decomposition Lemma]  \label{lemma:lower_dec}
Let ${\bf X} =  \{ X^n \}_{n=1}^\infty$ be a mixed memoryless source and $\overline{X} = \left\{ \overline{X}^n \right\}_{n=1}^\infty$ be an \textit{arbitrary} general source. Then, for any $\theta \in \Theta$, $z_n$ and $\gamma>0$ it holds that
\begin{IEEEeqnarray}{rCl}
\Pr\left\{  \frac{1}{n} \log \frac{P_{{X}^n}(X_\theta^n)}{P_{\overline{X}^n}(X_\theta^n)} \le z_n \right\} \ge \Pr\left\{ \frac{1}{n} \log \frac{P_{{X}_\theta^n}(X_\theta^n)}{P_{\overline{X}^n}(X_\theta^n)} \le z_n  - \frac{\gamma}{\sqrt{n}}  \right\} - e^{-\sqrt{n}\gamma}. 
\end{IEEEeqnarray}
\end{lemma}
\begin{IEEEproof}

Setting $\gamma> 0$, we define a set
\begin{IEEEeqnarray*}{rCl}
D_n & = & \left\{ {\bf x} \in {\cal X}^n \left| \frac{1}{n}\log P_{X_\theta^n}({\bf x}) - \frac{1}{n} \log P_{X^n}({\bf x})  \le - \frac{\gamma}{\sqrt{n}} \right.   \right\} ,
\end{IEEEeqnarray*}
for $\theta \in \Theta$.
Then, it holds that
\begin{IEEEeqnarray*}{rCl}
\Pr\left\{ X^n_\theta \in D_n \right\}  & = &  \sum_{{\bf x} \in D_n} P_{X_\theta^n}({\bf x}) \\
& \le & \sum_{{\bf x} \in D_n} P_{X^n}({\bf x})  e^{-\sqrt{n}\gamma} \\
& \le & e^{-\sqrt{n}\gamma}.
\end{IEEEeqnarray*}
Thus, for any {real number $z_n$} it holds that
\begin{IEEEeqnarray*}{rCl}
\lefteqn{\Pr\left\{   \frac{1}{n} \log  {P_{X_\theta^n}(X^n_\theta)} \le z_n -\frac{\gamma}{\sqrt{n}} \right\}}  \\
& = & \Pr\left\{   \frac{1}{n} \log  {P_{X_\theta^n}(X^n_\theta)} \le z_n -\frac{\gamma}{\sqrt{n}}, X_\theta^n \notin D_n \right\}  + \Pr\left\{   \frac{1}{n} \log  {P_{X_\theta^n}(X^n_\theta)} \le z_n -\frac{\gamma}{\sqrt{n}}, X_\theta^n \in D_n \right\} \\
& \le & \Pr\left\{   \frac{1}{n} \log  {P_{X^n}(X^n_\theta)} \le z_n \right\}  + \Pr\left\{ X_\theta^n \in D_n \right\} \\
& \le & \Pr\left\{   \frac{1}{n} \log  {P_{X^n}(X^n_\theta)} \le z_n \right\}  +  e^{-\sqrt{n}\gamma}.
\end{IEEEeqnarray*}
Hence, we obtain the inequality
\begin{IEEEeqnarray*}{rCl}
\Pr\left\{   \frac{1}{n} \log  {P_{X^n}(X^n_\theta)} \le z_n \right\}  
& \ge & \Pr\left\{   \frac{1}{n} \log  {P_{X_\theta^n}(X^n_\theta)} \le z_n -\frac{\gamma}{\sqrt{n}} \right\} -  e^{-\sqrt{n}\gamma},
\end{IEEEeqnarray*}
from which with $z_n + \frac{1}{n}\log P_{\overline{X}^n}(X_\theta^n)$ instead of $z_n$ it follows that
\begin{IEEEeqnarray*}{rCl}
\Pr\left\{   \frac{1}{n} \log \frac{P_{X^n}(X_\theta^n)}{P_{\overline{X}^n}(X_\theta^n)} \le z_n \right\}  
& \ge & \Pr\left\{  \frac{1}{n} \log \frac{P_{X_\theta^n}(X_\theta^n)}{P_{\overline{X}^n}(X_\theta^n)} \le z_n - \frac{\gamma}{\sqrt{n}}   \right\} -  e^{-\sqrt{n}\gamma} \\
\end{IEEEeqnarray*}
for all $\theta \in \Theta$.
This completes the proof. 
\end{IEEEproof}

The first-order $\varepsilon$-optimum exponent has been derived by using these three lemmas.
\begin{theorem}[First-order $\varepsilon$-optimum exponent: Han {\cite{Han}}] \label{theo:mixed_1st}
For $0 \le \varepsilon <1$,
\begin{equation}
B_{\varepsilon}({\bf X}||\overline{\bf X}) = \sup \left\{ R \left|  \int_{\{\theta|D(P_{X_\theta}||P_{\overline{X}}) < R\}} dw(\theta) \le \varepsilon \right. \right\} 
\end{equation}
where $D(P_{X}||P_{\overline{X}})$ denotes the divergence between $P_X$ and $P_{\overline{X}}$. 
\end{theorem}
\begin{IEEEproof}

As for the proof, see \cite{Han}.
\end{IEEEproof}
\begin{remark} \label{remark:singleton}
If $\Theta$ is a singleton, the above formula reduces to 
\[
B_\varepsilon({\bf X}||\overline{\bf X}) = D(P_X || P_{\overline{X}}) \quad (0\le \forall\varepsilon <1),
\]
which is nothing but Stein's lemma \cite{DZ}. 
\IEEEQED
\end{remark}
\begin{remark}  \label{remark:3-2}
$B_{\varepsilon}({\bf X}||\overline{\bf X})$ can be expressed also as 
\[
B_{\varepsilon}({\bf X}||\overline{\bf X}) = \sup \left\{ R \left|  \int_{\{\theta|D(P_{X_\theta}||P_{\overline{X}}) \le R\}} dw(\theta) \le \varepsilon \right. \right\}.
\]
This can be verified as follows.
Set 
\begin{IEEEeqnarray*}{rCl}
\tilde\beta_{\varepsilon} & := & \sup \left\{ R \left|  \int_{\{\theta|D(P_{X_\theta}||P_{\overline{X}}) \le R\}} dw(\theta) \le \varepsilon \right. \right\}, \\
{\beta}_{\varepsilon} & := & \sup \left\{ R \left|  \int_{\{\theta|D(P_{X_\theta}||P_{\overline{X}}) < R\}} dw(\theta) \le \varepsilon \right. \right\}.
\end{IEEEeqnarray*}
Then, clearly $\tilde{\beta}_{\varepsilon} \le {\beta}_{\varepsilon}$.
Here, we assume that $\tilde{\beta}_{\varepsilon} < {\beta}_{\varepsilon}$ to show a contradiction.
From the assumption, there exists a constant $\gamma >0$ satisfying
$\tilde{\beta}_{\varepsilon} + 2\gamma < {\beta}_{\varepsilon}$.
On the other hand, from the definition of $\beta_\varepsilon$, for any $\eta >0$
\begin{align*}
\varepsilon & \ge  \int_{\{\theta|D(P_{X_\theta}||P_{\overline{X}})< \beta_\varepsilon - \eta\}} dw(\theta) 
\end{align*}
holds. Thus, setting $\eta< \gamma$ leads to 
\begin{align*}
\varepsilon & \ge  \int_{\{\theta|D(P_{X_\theta}||P_{\overline{X}}) < \beta_\varepsilon - \eta\}} dw(\theta) \\
& \ge  \int_{\{\theta|D(P_{X_\theta}||P_{\overline{X}}) < \beta_\varepsilon - \gamma \}} dw(\theta) \\
& \ge  \int_{\{\theta|D(P_{X_\theta}||P_{\overline{X}}) \le \tilde{\beta}_\varepsilon +\gamma \}} dw(\theta) \\& > \varepsilon,
\end{align*}
which is a contradiction, where the last inequality is due to the definition of $\tilde{\beta}_\varepsilon$.
 \IEEEQED
\end{remark}
\subsection{Second-order $\varepsilon$-optimum exponent}
Next, we derive the second-order $\varepsilon$-optimum exponent for mixed sources.
\begin{theorem}[Second-order $\varepsilon$-optimum exponent: Han \cite{Han_Jeju}] \label{theo:mixed_2nd}
For $0 \le \varepsilon <1$, 
\begin{equation}
B_{\varepsilon}(R|{\bf X}||\overline{\bf X}) = \sup \left\{ S \left|  \int_{\{\theta| D(P_{X_\theta}||P_{\overline{X}}) <R \}} dw(\theta) + \int_{ \{\theta| D(P_{X_\theta}||P_{\overline{X}}) = R \} } \Phi_\theta(S) dw(\theta) \le \varepsilon \right. \right\},
\end{equation}
where 
\[
\Phi_\theta(S) := G\left( \frac{S}{\sqrt{V_\theta}} \right),
\]
\[
G(x) := \frac{1}{\sqrt{2\pi}}\int_{-\infty}^x e^{-\frac{x^2}{2}} dx,
\]
\[
V_\theta := \sum_{x \in {\cal X}} P_{X_\theta}(x) \left( \log \frac{P_{X_\theta}(x)}{P_{\overline{X}}(x)} - D(P_{X_\theta}||P_{\overline{X}}) \right)^2.
\]
\end{theorem}
\begin{remark}
If $\Theta$ is a singleton $(\Theta = \{\theta_0 \})$, Theorem \ref{theo:mixed_2nd} reduces to $B_\varepsilon(R|{\bf X}||\overline{\bf X}) = \sqrt{V_{\theta_0}}\Phi^{-1}_{\theta_0}(\varepsilon)$ for $R = B_\varepsilon({\bf X}||\overline{\bf X})$, which is originally due to Strassen \cite{Strassen}.
\end{remark}
\begin{IEEEproof}[Proof of Theorem \ref{theo:mixed_2nd}]

Setting
\[
\overline{B}_\varepsilon(R,S) := \int_{\{\theta| D(P_{X_\theta}||P_{\overline{X}}) <R \}} dw(\theta) + \int_{ \{\theta| D(P_{X_\theta}||P_{\overline{X}})= R \} } \Phi_\theta(S) dw(\theta),
\]
it suffices, in view of Theorem \ref{theo:general_2nd}, to show two inequalities:
\begin{IEEEeqnarray}{rCl} \label{eq:3-2-0-0}
 \overline{B}_\varepsilon(R,S) \ge \limsup_{n \to \infty} \Pr\left\{ \frac{1}{n}\log \frac{P_{X^n}(X^n)}{P_{\overline{X}^n}(X^n)} \le R + \frac{S}{\sqrt{n}} \right\} ,
\end{IEEEeqnarray}
\begin{IEEEeqnarray}{rCl}  \label{eq:3-2-0-1}
 \overline{B}_\varepsilon(R,S) \le \limsup_{n \to \infty} \Pr\left\{ \frac{1}{n}\log \frac{P_{X^n}(X^n)}{P_{\overline{X}^n}(X^n)} \le R + \frac{S}{\sqrt{n}} \right\}.
\end{IEEEeqnarray}

\textit{Proof of (\ref{eq:3-2-0-0})}:

By the definitions of ${\bf X}$ and ${\bf \overline{X}}$, it holds that
\begin{IEEEeqnarray}{rCl}  \label{eq:3-2-0}
\lefteqn{\limsup_{n \to \infty} \Pr\left\{ \frac{1}{n}\log \frac{P_{X^n}(X^n)}{P_{\overline{X}^n}(X^n)} \le  R + \frac{S}{\sqrt{n}}  \right\}} \nonumber  \\
& = &  \limsup_{n \to \infty} \int_{\Theta} dw(\theta)\Pr\left\{ \frac{1}{n}\log \frac{P_{X^n}(X_\theta^n)}{P_{\overline{X}^n}(X_\theta^n)} \le R + \frac{S}{\sqrt{n}}\right\} \nonumber \\
& \le & \limsup_{n \to \infty} \int_{\Theta_n^\ast} dw(\theta)\Pr\left\{ \frac{1}{n}\log \frac{P_{X^n}(X_\theta^n)}{P_{\overline{X}^n}(X_\theta^n)} \le R + \frac{S}{\sqrt{n}}\right\} + \limsup_{n \to \infty} \int_{\Theta-\Theta^\ast_n} dw(\theta) \nonumber \\
& = & \limsup_{n \to \infty} \int_{\Theta_n^\ast} dw(\theta)\Pr\left\{ \frac{1}{n}\log \frac{P_{X^n}(X_\theta^n)}{P_{\overline{X}^n}(X_\theta^n)} \le R + \frac{S}{\sqrt{n}}\right\} \nonumber \\
& \le & \limsup_{n \to \infty} \int_{\Theta_n^\ast} dw(\theta)\Pr\left\{ \frac{1}{n}\log \frac{P_{X_\theta^n}(X_\theta^n)}{P_{\overline{X}^n}(X_\theta^n)} \le  R + \frac{S}{\sqrt{n}}  + \frac{1}{\sqrt[4]{n}^3} \right\} \nonumber \\
& \le & \limsup_{n \to \infty} \int_{\Theta} dw(\theta)\Pr\left\{ \frac{1}{n}\log \frac{P_{X_\theta^n}(X_\theta^n)}{P_{\overline{X}^n}(X_\theta^n)} \le R + \frac{S}{\sqrt{n}}  + \frac{1}{\sqrt[4]{n}^3} \right\} \nonumber \\
& \le & \int_{\Theta} dw(\theta) \limsup_{n \to \infty} \Pr\left\{ \frac{1}{n}\log \frac{P_{X_\theta^n}(X_\theta^n)}{P_{\overline{X}^n}(X_\theta^n)} \le  R + \frac{S}{\sqrt{n}}  + \frac{1}{\sqrt[4]{n}^3} \right\},
\end{IEEEeqnarray}
where the second equality and the second inequality are due to Lemma \ref{lemma:expurgate} and  Lemma \ref{lemma:upper_dec}, respectively, and the last inequality is from Fatou's lemma.

Here, we define three sets:
\begin{IEEEeqnarray}{rCl}  \label{eq:three_sets}
\left.
\begin{array}{c}
\Theta_0 := \left\{ \theta \in \Theta  \left| D(P_{X_\theta}||P_{\overline{X}}) =  R \right. \right\},  \\
\Theta_1 := \left\{ \theta \in \Theta  \left|  D(P_{X_\theta}||P_{\overline{X}}) <  R \right. \right\},  \\ 
\Theta_2 := \left\{ \theta \in \Theta  \left| D(P_{X_\theta}||P_{\overline{X}}) > R \right. \right\}.
\end{array}
\right\}
\end{IEEEeqnarray}
Noting that, setting $X_\theta^n = (X_{\theta,1},X_{\theta,2},\cdots, X_{\theta,n})$,
\begin{IEEEeqnarray}{rCl} \label{eq:3-0-0-1}
\frac{1}{n}\log \frac{P_{X_\theta^n}(X^n_\theta)}{P_{\overline{X}^n}(X^n_\theta)} = \frac{1}{n} \sum_{i=1}^n \log \frac{P_{X_\theta}(X_{\theta,i})}{P_{\overline{X}}(X_{\theta,i})}
\end{IEEEeqnarray}
gives the arithmetic average of $n$ i.i.d. variables with expectation
\begin{IEEEeqnarray}{rCl} \label{eq:3-0-0-2}
\mathbb{E} \left[ \frac{1}{n} \sum_{i=1}^n \log \frac{P_{X_\theta}(X_\theta)}{P_{\overline{X}}(X_\theta)} \right]
 = D(P_{X_\theta}||P_{\overline{X}}).
\end{IEEEeqnarray}
Then, the weak law of large numbers yields that for $\forall \theta \in \Theta_2$
\begin{equation}  \label{eq:law}
\limsup_{n \to \infty} \Pr\left\{ \frac{1}{n}\log \frac{P_{X_\theta^n}(X_\theta^n)}{P_{\overline{X}^n}(X_\theta^n)}  \le R + \frac{S}{\sqrt{n}}  + \frac{1}{\sqrt[4]{n}^3} \right\} = 0. 
\end{equation}
Moreover, for $\forall \theta \in \Theta_0$ the central limit theorem leads to
\begin{eqnarray}  \label{eq:clt}
\lefteqn{\limsup_{n \to \infty} \Pr\left\{ \frac{1}{n}\log \frac{P_{X_\theta^n}(X_\theta^n)}{P_{\overline{X}^n}(X_\theta^n)} \le R + \frac{S}{\sqrt{n}}  + \frac{1}{\sqrt[4]{n}^3} \right\}} \nonumber \\
& = & \limsup_{n \to \infty} \Pr\left\{ \frac{1}{\sqrt{n}}\left(\log \frac{P_{X_\theta^n}(X_\theta^n)}{P_{\overline{X}^n}(X_\theta^n)} - \sqrt{n} D(P_{X_\theta}||P_{\overline{X}}) \right) \le S +   \frac{1}{\sqrt[4]{n}} \right\} \nonumber \\
&  = & \Phi_\theta \left(S \right).
\end{eqnarray}

Summarizing these equalities, we obtain
\begin{eqnarray} \label{eq:3-2-1}
\lefteqn{\int_{\Theta} dw(\theta) \limsup_{n \to \infty} \Pr\left\{ \frac{1}{n}\log \frac{P_{X_\theta^n}(X_\theta^n)}{P_{\overline{X}^n}(X_\theta^n)} \le  R + \frac{S}{\sqrt{n}}  + \frac{1}{\sqrt[4]{n}^3} \right\}} \nonumber  \\
 & = &  \int_{\Theta_1} dw(\theta) \limsup_{n \to \infty} \Pr\left\{ \frac{1}{n}\log \frac{P_{X_\theta^n}(X_\theta^n)}{P_{\overline{X}^n}(X_\theta^n)} \le R + \frac{S}{\sqrt{n}}  + \frac{1}{\sqrt[4]{n}^3} \right\} \nonumber \\
 && + \int_{\Theta_0} dw(\theta) \limsup_{n \to \infty} \Pr\left\{ \frac{1}{n}\log \frac{P_{X_\theta^n}(X_\theta^n)}{P_{\overline{X}^n}(X_\theta^n)} \le R + \frac{S}{\sqrt{n}}  + \frac{1}{\sqrt[4]{n}^3} \right\} \nonumber \\
 & \le & \int_{\{\theta | D(P_{X_\theta}||P_{\overline{X}}) < R \} } dw(\theta)  + \int_{\{\theta | D(P_{X_\theta}||P_{\overline{X}}) = R \} } \Phi_\theta \left(S \right)  dw(\theta).
\end{eqnarray}
Plugging (\ref{eq:3-2-1}) into (\ref{eq:3-2-0}) yields (\ref{eq:3-2-0-0}).

\textit{Proof of (\ref{eq:3-2-0-1})}:

By definitions of ${\bf X}$ and ${\bf \overline{X}}$, and Lemma \ref{lemma:lower_dec} with $z_n = R + \frac{S}{\sqrt{n}}$, it holds that
\begin{IEEEeqnarray}{rCl}  \label{eq:3-2-3}
\lefteqn{\limsup_{n \to \infty} \Pr\left\{ \frac{1}{n}\log \frac{P_{X^n}(X^n)}{P_{\overline{X}^n}(X^n)} \le R + \frac{S}{\sqrt{n}} \right\}} \nonumber \\
 & \ge & \liminf_{n \to \infty} \int_{\Theta} dw(\theta)\Pr\left\{ \frac{1}{n}\log \frac{P_{X^n}(X_\theta^n)}{P_{\overline{X}^n}(X_\theta^n)} \le R + \frac{S}{\sqrt{n}} \right\} \nonumber \\
& \ge & \liminf_{n \to \infty} \int_{\Theta} dw(\theta)\Pr\left\{ \frac{1}{n}\log \frac{P_{X_\theta^n}(X_\theta^n)}{P_{\overline{X}^n}(X_\theta^n)} \le R + \frac{S}{\sqrt{n}} - \frac{\gamma}{\sqrt{n}}\right\} \nonumber \\
& = & \liminf_{n \to \infty} \int_{\Theta} dw(\theta)\Pr\left\{ \frac{1}{n}\log \frac{P_{X_\theta^n}(X_\theta^n)}{P_{\overline{X}^n}(X_\theta^n)} \le R + \frac{S-\gamma}{\sqrt{n}} \right\} \nonumber \\
& \ge & \int_{\Theta} dw(\theta) \liminf_{n \to \infty} \Pr\left\{ \frac{1}{n}\log \frac{P_{X_\theta^n}(X_\theta^n)}{P_{\overline{X}^n}(X_\theta^n)} \le R + \frac{S-\gamma}{\sqrt{n}} \right\},
\end{IEEEeqnarray}
for any $\gamma >0$. We also partition the parameter space $\Theta$ into three sets as in (\ref{eq:three_sets}) in the proof of (\ref{eq:3-2-0-0}).

Then, similarly to the derivation of (\ref{eq:law}) and (\ref{eq:clt}), we obtain
\begin{equation} \label{eq:law2}
\liminf_{n \to \infty} \Pr\left\{ \frac{1}{n}\log \frac{P_{X_\theta^n}(X_\theta^n)}{P_{\overline{X}^n}(X_\theta^n)}  \le R + \frac{S -\gamma}{\sqrt{n}} \right\} =  \left\{ \begin{array}{cc}%
\Phi_\theta\left(S -\gamma  \right), \quad & \theta \in \Theta_0 \\
1.  \quad & \theta \in \Theta_1 
\end{array}\right.
\end{equation}
Thus, the right-hand side of (\ref{eq:3-2-3}) is rewritten as
\begin{IEEEeqnarray}{rCl} \label{eq:3-2-3-1}
\lefteqn{\int_{\Theta} dw(\theta) \liminf_{n \to \infty} \Pr\left\{ \frac{1}{n}\log \frac{P_{X_\theta^n}(X_\theta^n)}{P_{\overline{X}^n}(X_\theta^n)} \le R + \frac{S-\gamma}{\sqrt{n}}  \right\}} \nonumber \\
& \ge & \int_{\Theta_1} dw(\theta) \liminf_{n \to \infty} \Pr\left\{ \frac{1}{n}\log \frac{P_{X_\theta^n}(X_\theta^n)}{P_{\overline{X}^n}(X_\theta^n)} \le R + \frac{S-\gamma}{\sqrt{n}}\right\} \nonumber \\
& & + \int_{\Theta_0} dw(\theta) \liminf_{n \to \infty} \Pr\left\{ \frac{1}{n}\log \frac{P_{X_\theta^n}(X_\theta^n)}{P_{\overline{X}^n}(X_\theta^n)} \le R + \frac{S-\gamma}{\sqrt{n}}\right\} \nonumber \\
 & = & \int_{\{\theta | D(P_{X_\theta}||P_{\overline{X}}) < R  \} } dw(\theta) + \int_{\{\theta | D(P_{X_\theta}||P_{\overline{X}}) = R  \} }  \Phi_\theta \left( S - \gamma \right) dw(\theta). 
\end{IEEEeqnarray}
Substituting (\ref{eq:3-2-3-1}) into (\ref{eq:3-2-3}) and noting that $\gamma >0$ is arbitrary, we obtain (\ref{eq:3-2-0-1}).
\end{IEEEproof}
\begin{remark}
From Theorem \ref{theo:mixed_1st} with $R=B_\varepsilon({\bf X}||\overline{\bf X})$, it is not difficult to verify that
\begin{equation} \label{eq:c-1}
\int_{\{\theta | D(P_{X_\theta}||P_{\overline{X}}) < R  \} } dw(\theta) \le \varepsilon
\end{equation}
and
\begin{equation}  \label{eq:c-2}
\int_{\{\theta | D(P_{X_\theta}||P_{\overline{X}}) \le R  \} } dw(\theta) \ge \varepsilon.
\end{equation}
Here, let us consider the following canonical equation for $S$
\begin{equation}  \label{eq:c-3}
\int_{\Theta } dw(\theta) \lim_{n \to \infty} \Phi_\theta(\sqrt{n} (B_\varepsilon({\bf X}||\overline{\bf X}) - D(P_{X_\theta}||P_{\overline{X}})) + S) = \varepsilon.
\end{equation}
In view of (\ref{eq:c-1}) and (\ref{eq:c-2}), this equation always has a solution $S = S(\varepsilon)$.
It should be noted that if $\int_{\{\theta | D(P_{X_\theta}||P_{\overline{X}}) = B_\varepsilon({\bf X}||\overline{\bf X})  \} } dw(\theta) = 0 $ holds, the solution is not unique and so $S(\varepsilon) = +\infty$.
By using the solution $S(\varepsilon)$, it is not difficult to check that Theorem \ref{theo:mixed_2nd} with $R=B_\varepsilon({\bf X}||\overline{\bf X})$ can be expressed as 
\[
B_\varepsilon(R|{\bf X}||\overline{\bf X}) = S(\varepsilon).
\]
The canonical equation is a useful expression for the \textit{second-order} $\varepsilon$-optimum rate \cite{PPV2011,NH2011,NH2014,YHN2016}.
The equation (\ref{eq:c-3}) is the hypothesis testing counterpart of these results.
\IEEEQED
\end{remark}

%
%
%
%
\section{Mixed memoryless alternative hypothesis}
In this section, we consider the case where not only the null hypothesis but also the alternative hypothesis is a mixed memoryless source. 

Let $\left\{{P}_{\overline{X}_\sigma}\right\}_{\sigma \in \Sigma}$ be a family of probability distributions on ${\cal X}$ where $\Sigma$ is a probability space with probability measure $v(\sigma)$.
We assume here that $\Sigma$ is a \textit{compact} space and ${P}_{\overline{X}_\sigma}$ is \textit{continuous} as a function of $\sigma \in \Sigma$.

The hypothesis testing problem considered in this section is stated as follows.
\begin{itemize}
\item The null hypothesis is a mixed memoryless source ${\bf X} = \{X^n \}_{n=1}^\infty$, that is, for $\forall {\bf x} \in  {\cal X}^n$ 
\begin{equation} \label{eq:mix2}
P_{X^n}({\bf x}) = \int_\Theta P_{X^n_\theta}({\bf x}) dw(\theta),
\end{equation}
where
\[
P_{X^n_\theta}({\bf x}) = \prod_{i=1}^n P_{X_\theta}(x_i).
\]
\item The alternative hypothesis is another mixed memoryless source $\overline{\bf X}= \left\{\overline{X}^n \right\}_{n=1}^\infty$, that is, for $\forall {\bf x} \in  {\cal X}^n$ 
\begin{equation} \label{eq:mix3}
P_{\overline{X}^n}({\bf x}) = \int_{{\Sigma}} P_{\overline{X}^n_{\sigma}}({\bf x}) dv(\sigma),
\end{equation}
where 
\[
P_{\overline{X}^n_{\sigma}}({\bf x}) = \prod_{i=1}^n P_{\overline{X}_{{\sigma}}}(x_i).
\]
\end{itemize}
For simplicity, we may write $P_\theta, {P}^n_\theta$ (resp. $\overline{P}_\sigma,\overline{P}^n_\sigma$) instead of ${P}_{X_\theta}, {P}_{X^n_\theta}$ (resp. ${P}_{\overline{X}_\sigma}, {P}_{\overline{X}^n_\sigma}$).
We assume also that ${|\cal X|} < \infty$. 

\begin{theorem}[First order $\varepsilon$-optimum exponent] \label{theo:mixed2_1st}
For $0 \le \varepsilon <1$,
\begin{equation}  \label{eq:theo:mixed2_1st}
B_{\varepsilon}({\bf X}||\overline{\bf X}) = \sup \left\{ R \left|  \int_{\{\theta|D(P_\theta||\overline{P}_{\sigma(P_\theta)}) < R\}} dw(\theta) \le \varepsilon \right. \right\},
\end{equation}
where the function $\sigma(P)$ is specified by the equation
\begin{IEEEeqnarray}{rCl} \label{eq:sigma}
D(P||\overline{P}_{\sigma(P)})  =  v\mbox{-ess.} \inf D(P||\overline{P}_\sigma)
\end{IEEEeqnarray}
 and $ v\mbox{-ess.} \inf f_\sigma := \sup\{ \beta |\Pr \{ f_\sigma < \beta \} =0 \}$ (the essential infimum of $f_\sigma$ with respect to $v(\sigma)$; $\lq\lq \Pr"$ is measured with respect to the probability measure $v(\sigma)$).
\end{theorem}
\begin{remark}
Notice here that $D(P||\overline{P}_\sigma)$ is continuous in $(P, \overline{P}_\sigma)$.
Since we have assumed that $\Sigma$ is compact and $\overline{P}_\sigma$ is continuous in $\sigma$, there indeed exists a continuous function $\sigma(P)$ satisfying (\ref{eq:sigma}).
\IEEEQED
\end{remark}
\begin{remark}
In the case that $\Sigma$ is a singleton, the above theorem coincides with Theorem \ref{theo:mixed_1st}.
Therefore, this theorem is a direct generalization of Theorem \ref{theo:mixed_1st}. 
This means that if both $\Theta$ and $\Sigma$ are singletons the theorem coincides with Stein's lemma.
(See, Remark \ref{remark:singleton}.)
\IEEEQED
\end{remark}
\begin{remark}  \label{remark:4-2}
Remark \ref{remark:3-2} is also valid in this theorem. That is, $B_\varepsilon({\bf X}||\overline{\bf X})$ is expressed also as
\begin{equation}
B_{\varepsilon}({\bf X}||\overline{\bf X}) = \sup \left\{ R \left|  \int_{\{\theta|D(P_\theta||\overline{P}_{\sigma(P_\theta)}) \le R\}} dw(\theta) \le \varepsilon \right. \right\}.
\end{equation} \IEEEQED
\end{remark}

\begin{IEEEproof}[Proof of Theorem \ref{theo:mixed2_1st}]

In order to show the theorem, let ${\cal T}_{\theta,\nu}^n  \subseteq {\cal X}^n$ be the set of $\nu$-typical sequence  with respect to $P_{X_\theta}$, that is, let ${\cal T}_{\theta,\nu}^n$ be the set of all ${\bf x}=(x_1,x_2,\cdots,x_n) \in {\cal X}^n$ such that
\[
\left| N(x |{\bf x})/n - {P_{X_\theta}({x})} \right| \le \nu P_{X_\theta}(x) \quad (\forall x \in {\cal X}),
\]
where $N(x |{\bf x})$ is the number of $i$ such that $x_i=x$, and $\nu >0$ is an arbitrary constant.
Then, it is well known that 
\begin{equation} \label{eq:typical}
\Pr\left\{ X_\theta^n \in {\cal T}_{\theta,\nu}^n \right\} \to 1 \quad (n \to \infty).
\end{equation}
We first derive the upper and lower bounds for the probability
\begin{align} \label{eq:Ax}
P_{\overline{X}^n}({\bf x}) = \int_{{\Sigma}} \overline{P}^n_\sigma({\bf x}) dv(\sigma),
\end{align}
for any fixed ${\bf x} \in {\cal T}_{\theta,\nu}^n$.

In order to upper bound (\ref{eq:Ax}), we define $a({\bf x})$ as
\begin{align*} 
a({\bf x}) := v\mbox{-ess.} \sup \overline{P}^n_\sigma({\bf x}),
\end{align*}
where $ v\mbox{-ess.} \sup f_\sigma$ denotes the essential supremum of $f_\sigma$ with respect to $v(\sigma)$, i.e.,
$ v\mbox{-ess.} \sup f_\sigma := \inf\{ \alpha| \Pr\{ f_\sigma>\alpha\}=0 \}$.
Thus, from the property of the essential supremum we  immediately have
\begin{align}  \label{eq:4-l-1}
a({\bf x})  \ge P_{\overline{X}^n}({\bf x}),
\end{align}
for $\forall n=1,2,\cdots$.

Let $P_{\bf x}$ denote the type of ${\bf x} \in {\cal T}_{\theta,\nu}^n $. Then, noting that
\begin{align*}
\overline{P}^n_\sigma({\bf x}) & = \sum_{x \in {\cal X}} \overline{P}_\sigma(x)^{N(x|{\bf x})} \nonumber \\
& = \exp\left[\sum_{x \in {\cal X}} {N(x|{\bf x})} \log \overline{P}_\sigma(x)  \right] \nonumber \\
& = \exp\left[ -n\left( H(P_{\bf x}) + D(P_{\bf x}||\overline{P}_\sigma) \right)\right]
\end{align*}
holds,
$a({\bf x})$ is written as 
\[
a({\bf x}) = \exp\left[ -n\left( H(P_{\bf x}) + v\mbox{-ess.} \inf D(P_{\bf x}||\overline{P}_\sigma) \right)\right].
\]

Here, it is important to notice that $D(P||\overline{P}_\sigma)$ is continuous in $(P,\overline{P}_\sigma)$ owing to the assumption and hence $D(P||\overline{P}_{\sigma(P)})$ is \textit{continuous} in $P \in {\cal P}({\cal X})$ (the set of probability distributions on ${\cal X}$).
Thus, expanding $D(P||\overline{P}_{\sigma(P_{\bf x})})$ in $P_{\bf x}$ around  $P_\theta$ leads to
\begin{align*}
D\left(P||\overline{P}_{\sigma(P_{\bf x})}\right) & = D(P||\overline{P}_{\sigma(P_{\theta})}) + \delta_\theta(\nu) \quad ({\bf x} \in {\cal T}_{\theta,\nu}^n ).
\end{align*}
where $\delta_\theta(\nu) \to 0$ as $\nu \to 0$,
because $\sum_{x \in {\cal X}}  |P_\theta(x) - P_{\bf x}(x)| \le \nu$ for ${\bf x} \in {\cal T}_{\theta,\nu}^n$ .

Then, in view of (\ref{eq:4-l-1}) for each ${\bf x} \in  {\cal T}_{\theta,\nu}^n$ we have the upper bound:
\begin{align}
{P}_{\overline{X}^n}({\bf x}) & \le  a({\bf x}) \nonumber \\
& = \exp\left[ -n\left( H(P_{\bf x}) + v\mbox{-ess.} \inf D(P_{\bf x}||\overline{P}_{\sigma(P_{\bf x})}) \right)\right] \nonumber \\
& =\exp\left[ -n\left( H(P_{\bf x}) +  D(P_{\bf x}||\overline{P}_{\sigma(P_{\theta})}) - \delta_\theta(\nu) \right)\right] \nonumber \\
& =\overline{P}^n_{\sigma(P_\theta)}({\bf x}) \exp[n \delta_\theta(\nu)],
\end{align}
from which it follows that
\begin{align} \label{eq:upper2}
\frac{1}{n} \log \frac{1}{P_{\overline{X}^n}({\bf x}) } & \ge  \frac{1}{n} \log \frac{1}{\overline{P}^n_{\sigma(P_\theta)}({\bf x})} - \delta_\theta(\nu)
\end{align}
for each ${\bf x} \in  {\cal T}_{\theta,\nu}^n$.

Next, we show the lower bound of $P_{\overline{X}^n}({\bf x})$.
For any $P \in {\cal P}({\cal X})$ and any small constant $\tau >0$, set
\begin{align}  \label{eq:S_tau}
{\cal S}_\tau(P) : = \left\{ \sigma \in \Sigma \left|  D(P||\overline{P}_\sigma) < D(P||\overline{P}_{\sigma(P)}) + \tau  \right. \right\},
\end{align}
then, by the definition of $v$-ess.inf.,
\begin{align*}
c_\tau(P) := \int_{{\cal S}_\tau(P)} d v(\sigma) > 0
\end{align*}
holds.
Our claim is that for any $\theta \in \Theta$ and sufficiently small $\tau >0$ and with some positive constant $c_\theta>0$
\begin{equation}  \label{eq:claim}
\inf_{{\bf x} \in  {\cal T}_{\theta,\nu}^n}c_\tau(P_{\bf x}) \ge c_\theta.
\end{equation}
To see this, consider a sequence $\{\tau_i \}_{i=1}^\infty$ such that $ 0< \tau_1 < \tau_2 < \cdots \to \tau$.
Then, there exists a positive integer $m$ such that $c_{\tau_{m}}(P_{\theta}) > 0$.
Otherwise, the continuity of probability measure implies that 
\[
0 = \lim_{i \to \infty} c_{\tau_i}(P_{\theta})  =c_{\tau}(P_{\theta}) >0,
\]
which is a contradiction.
On the other hand, in view of (\ref{eq:S_tau}), $\sigma \in {\cal S}_{\tau_{m}}(P_\theta)$ is equivalent to 
\begin{align}\label{eq:4-5-1}
D(P_\theta||\overline{P}_\sigma) < D(P_\theta||\overline{P}_{\sigma(P_\theta)}) + \tau_{m}.
\end{align}
Since $D(P||\overline{P}_\sigma)$ and $D(P||\overline{P}_{\sigma(P)})$ are continuous in $P \in {\cal P}({\cal X})$ around $P = P_\theta$, if $\nu >0$ is sufficiently small then from (\ref{eq:4-5-1}) it follows that 
\begin{align}\label{eq:4-5-2}
D(P_{{\bf x}}||\overline{P}_\sigma) < D(P_{\bf x}||\overline{P}_{\sigma(P_{\bf x})}) + \tau_{m} + \gamma(\nu) \quad (\forall {\bf x} \in {\cal T}_{\theta,\nu}^n),
\end{align}
where we also have used the expansion in $P_{\bf x}$ around $P_\theta$ and $\gamma(\nu) \to 0$ as $\nu \to 0$.
Therefore, all $\sigma \in {\cal S}_{\tau_{m}}(P_\theta)$ satisfy (\ref{eq:4-5-2}).
Now we can take $\nu >0$ so that $\tau_{m} + \gamma(\nu) < \tau$ to have 
\begin{align}\label{eq:4-5-3}
D(P_{\bf x}||\overline{P}_\sigma) < D(P_{\bf x}||\overline{P}_{\sigma(P_{\bf x})}) + \tau \quad (\forall {\bf x} \in {\cal T}_{\theta,\nu}^n).
\end{align}
Therefore, ${\cal S}_{\tau_{m}}(P_\theta) \subset {\cal S}_{\tau}(P_{\bf x})$.
Hence, we have
\begin{equation}  \label{eq:4-5-4}
0 < c_{\tau_{m}}(P_{\theta}) \le c_{\tau}(P_{\bf x})\quad (\forall {\bf x} \in {\cal T}_{\theta,\nu}^n).
\end{equation}
This is nothing but (\ref{eq:claim}).

Thus, again for $\forall {\bf x} \in {\cal T}_{\theta,\nu}^n$ we have the lower bound
\begin{align} \label{eq:lower}
{P}_{\overline{X}^n}({\bf x}) & =   \int_{{\Sigma}} \overline{P}^n_\sigma({\bf x}) dv(\sigma)\nonumber \\
& \ge  \int_{{\cal S}_\tau(P_{\bf x})} \overline{P}^n_\sigma({\bf x}) dv(\sigma) \nonumber \\
& =  \int_{{\cal S}_\tau(P_{\bf x})}\exp\left[ -n\left( H(P_{\bf x}) + D(P_{\bf x}||\overline{P}_{\sigma}) \right)\right] dv(\sigma) \nonumber \\
& \ge  \int_{{\cal S}_\tau(P_{\bf x})}\exp\left[ -n\left( H(P_{\bf x}) + D(P_{\bf x}||\overline{P}_{\sigma(P_{\bf x})}) + \tau \right) \right] dv(\sigma) \nonumber \\
& = c_\tau(P_{\bf x}) \exp\left[ -n\left( H(P_{\bf x}) + D(P_{\bf x}||\overline{P}_{\sigma(P_{\theta})}) + \tau - \delta_\theta(\nu) \right) \right] \nonumber \\
& \ge c_{\tau_{m}}(P_{\theta}) \overline{P}^n_{\sigma(P_\theta)}({\bf x}) \exp\left[ n\left( \delta_\theta(\nu) - \tau \right) \right], 
\end{align}
where in the last equality and in the last inequality we have used the continuity of 
$D\left(P_{\bf x}||\overline{P}_{\sigma(P_{\bf x})}\right)$ in $P_{\bf x}$ around $P_\theta$ and (\ref{eq:4-5-4}), respectively.
From (\ref{eq:lower}), we obtain
\begin{align} \label{eq:lower2}
\frac{1}{n} \log \frac{1}{P_{\overline{X}^n}({\bf x}) } & \le  \frac{1}{n} \log \frac{1}{\overline{P}^n_{\sigma(P_\theta)}({\bf x})}  + \frac{1}{n} \log \frac{1}{c_{\tau_{m}}(P_{\theta})}    +   (\tau - \delta_\theta(\nu) ),
\end{align}
for each ${\bf x} \in  {\cal T}_{\theta,\nu}^n$.

We now turn to prove the theorem by using (\ref{eq:upper2}) and (\ref{eq:lower2}).
In view of Theorem \ref{theo:general_1st} and Remark \ref{remark:4-2}, it suffices to show two inequalities:
\begin{equation} \label{eq:4-0-0-1}
\limsup_{n \to \infty} \Pr\left\{ \frac{1}{n}\log \frac{P_{X^n}(X^n)}{P_{\overline{X}^n}(X^n)} \le R\right\}  \le   \int_{\{\theta|D(P_\theta||\overline{P}_{\sigma(P_\theta)}) \le R\}} dw(\theta),
\end{equation}
\begin{equation}  \label{eq:4-0-0-2}
\limsup_{n \to \infty} \Pr\left\{ \frac{1}{n}\log \frac{P_{X^n}(X^n)}{P_{\overline{X}^n}(X^n)} \le R\right\}  \ge   \int_{\{\theta|D(P_\theta||\overline{P}_{\sigma(P_\theta)})< R\}} dw(\theta).
\end{equation}

\textit{Proof of (\ref{eq:4-0-0-1})}:

Similarly to the derivation of (\ref{eq:3-2-0}) with  Lemma \ref{lemma:upper_dec}, we have
\begin{IEEEeqnarray}{rCl}  \label{eq:4-0}
\lefteqn{\limsup_{n \to \infty} \Pr\left\{ \frac{1}{n}\log \frac{P_{X^n}(X^n)}{P_{\overline{X}^n}(X^n)} \le R \right\}} \nonumber \\
 & = & \limsup_{n \to \infty} \int_{\Theta} dw(\theta)\Pr\left\{ \frac{1}{n}\log \frac{P_{X^n}(X_{\theta}^n)}{P_{\overline{X}^n}(X_{\theta}^n)} \le R \right\} \nonumber \\
& \le & \int_{\Theta} dw(\theta)  \limsup_{n \to \infty} \Pr\left\{ \frac{1}{n}\log \frac{P_{X_\theta^n}(X_{\theta}^n)}{P_{\overline{X}^n}(X_{\theta}^n)} \le R + \frac{1}{\sqrt[4]{n}^3} \right\}.
\end{IEEEeqnarray}
From the definition of the $\nu$-typical set and (\ref{eq:upper2}), we also have
\begin{IEEEeqnarray}{rCl}  \label{eq:4-0-0-3}
\lefteqn{\limsup_{n \to \infty} \Pr\left\{ \frac{1}{n}\log \frac{P_{X_\theta^n}(X_{\theta}^n)}{P_{\overline{X}^n}(X_{\theta}^n)} \le R + \frac{1}{\sqrt[4]{n}^3} \right\}} \nonumber \\
& \le & \limsup_{n \to \infty} \Pr\left\{ \frac{1}{n}\log \frac{P_{X_\theta^n}(X_{\theta}^n)}{P_{\overline{X}^n}(X_{\theta}^n)} \le R + \frac{1}{\sqrt[4]{n}^3}, \ X_\theta^n \in   {\cal T}_{\theta,\nu}^n\right\} + \limsup_{n \to \infty} \Pr\left\{ X_\theta^n \notin   {\cal T}_{\theta,\nu}^n\right\} \nonumber \\
& \le & \limsup_{n \to \infty} \Pr\left\{   \frac{1}{n}\log \frac{P_{X_\theta^n}(X_{\theta}^n)}{\overline{P}^n_{\sigma(P_\theta)}(X_{\theta}^n)} \le R + \frac{1}{\sqrt[4]{n}^3} + \delta_\theta(\nu), \ X_\theta^n \in   {\cal T}_{\theta,\nu}^n\right\}  \nonumber \\
& \le & \limsup_{n \to \infty} \Pr\left\{   \frac{1}{n}\log \frac{P_{X_\theta^n}(X_{\theta}^n)}{\overline{P}^n_{\sigma(P_\theta)}(X_{\theta}^n)} \le R + \frac{1}{\sqrt[4]{n}^3} + \delta_\theta(\nu)\right\},
\end{IEEEeqnarray}
for any $\theta \in \Theta$.
Here, we define two sets:
\begin{IEEEeqnarray*}{rCl}
\Theta_1 &:=& \left\{ \theta \in \Theta  \left| D(P_\theta||\overline{P}_{\sigma(P_\theta)}) \le  R  \right. \right\}, \\
\Theta_2 &:=& \left\{ \theta \in \Theta  \left| D(P_\theta||\overline{P}_{\sigma(P_\theta)})>  R  \right. \right\}.
\end{IEEEeqnarray*}
Then, from the definition of $\Theta_2$ there exists a small constant $\gamma >0$ satisfying
\[
D(P_\theta||\overline{P}_{\sigma(P_\theta)}) >  R +  3\gamma
\]
for $\theta \in \Theta_2$.
Thus, it holds that
\begin{eqnarray} \label{eq:4-5-5}
\lefteqn{\limsup_{n \to \infty} \Pr\left\{  \frac{1}{n}\log \frac{P_{X_\theta^n}(X_{\theta}^n)}{\overline{P}^n_{\sigma(P_\theta)}(X_{\theta}^n)} \le R + \frac{1}{\sqrt[4]{n}^3} + \delta_\theta(\nu) \right\}} \nonumber \\
& \le & \limsup_{n \to \infty} \Pr\left\{  \frac{1}{n}\log \frac{P_{X_\theta^n}(X_{\theta}^n)}{\overline{P}^n_{\sigma(P_\theta)}(X_{\theta}^n)} \le D(P_\theta||\overline{P}_{\sigma(P_\theta)}) + \frac{1}{\sqrt[4]{n}^3} + \delta_\theta(\nu) - 3\gamma \right\} \nonumber \\
& \le & \limsup_{n \to \infty} \Pr\left\{  \frac{1}{n}\log \frac{P_{X_\theta^n}(X_{\theta}^n)}{\overline{P}^n_{\sigma(P_\theta)}(X_{\theta}^n)} \le D(P_\theta||\overline{P}_{\sigma(P_\theta)}) - \gamma \right\},
\end{eqnarray}
where we have used the relation $\frac{1}{\sqrt[4]{n}^3} < \gamma$, and $\delta_\theta(\nu) < \gamma$
for sufficiently large $n$ and sufficiently small $\nu >0$.

Therefore, noting that, with $X_\theta^n = (X_{\theta,1},X_{\theta,2},\cdots,X_{\theta,n})$,  
\[
\frac{1}{n}\log \frac{P_{X_\theta^n}(X^n_\theta)}{\overline{P}^n_{\sigma(P_\theta)}(X^n_\theta)} = \frac{1}{n} \sum_{i=1}^n \log \frac{P_{X_\theta}(X_{\theta,i})}{\overline{P}_{\sigma(P_\theta)}(X_{\theta,i})}
\]
gives the arithmetic average of $n$ i.i.d. variables with expectation
\[
\mathbb{E} \left[ \frac{1}{n} \sum_{i=1}^n \log \frac{P_{X_\theta}(X_\theta)}{\overline{P}^n_{\sigma(P_\theta)}(X_\theta)} \right]
 = D(P_\theta||\overline{P}_{\sigma(P_\theta)}).
\]
 Then, the weak law of large numbers yields that for $\forall \theta \in \Theta_2$,
\begin{equation}\label{eq:4-5-6}
\limsup_{n \to \infty} \Pr\left\{  \frac{1}{n}\log \frac{P_{X_\theta^n}(X_{\theta}^n)}{\overline{P}^n_{\sigma(P_\theta)}(X_{\theta}^n)} \le  D(P_\theta||\overline{P}_{\sigma(P_\theta)}) - \gamma \right\} =  0.
\end{equation}
 Thus, from (\ref{eq:4-5-5}) and (\ref{eq:4-5-6}), the right-hand side of (\ref{eq:4-0}) is upper bounded by
\begin{eqnarray} \label{eq:4-1}
\lefteqn{ \int_{\Theta} dw(\theta)  \limsup_{n \to \infty} \Pr\left\{ \frac{1}{n}\log \frac{P_{X_\theta^n}(X_{\theta}^n)}{\overline{P}^n_{\sigma(P_\theta)}(X_{\theta}^n)} \le R + \frac{1}{\sqrt[4]{n}^3} + \delta_\theta(\nu) \right\}} \nonumber \\
 & \le &  \int_{\Theta_1} dw(\theta) \limsup_{n \to \infty} \Pr\left\{ \frac{1}{n}\log \frac{P_{X_\theta^n}(X_{\theta}^n)}{\overline{P}^n_{\sigma(P_\theta)}(X_{\theta}^n)} \le D(P_\theta||\overline{P}_{\sigma(P_\theta)}) -\gamma \right\} \nonumber \\
 && + \int_{\Theta_2} dw(\theta) \limsup_{n \to \infty} \Pr\left\{ \frac{1}{n}\log \frac{P_{X_\theta^n}(X_{\theta}^n)}{\overline{P}^n_{\sigma(P_\theta)}(X_{\theta}^n)} \le D(P_\theta||\overline{P}_{\sigma(P_\theta)}) -\gamma \right\} \nonumber \\
 & \le & \int_{\Theta_1} dw(\theta) \nonumber \\
 & = & \int_{\{\theta | D(P_\theta||\overline{P}_{\sigma(P_\theta)})\le  R  \} } dw(\theta),
\end{eqnarray}
which completes the proof of (\ref{eq:4-0-0-1}).

\textit{Proof of (\ref{eq:4-0-0-2}):}

Similarly to the derivation of (\ref{eq:3-2-3}) with Lemma \ref{lemma:lower_dec}, we have
\begin{IEEEeqnarray}{rCl}  \label{eq:4-3}
\lefteqn{\limsup_{n \to \infty} \Pr\left\{ \frac{1}{n}\log \frac{P_{X^n}(X^n)}{P_{\overline{X}^n}(X^n)} \le R \right\} } \nonumber \\
& \ge & \int_{\Theta} dw(\theta) \liminf_{n \to \infty} \Pr\left\{ \frac{1}{n}\log \frac{P_{X_\theta^n}(X_\theta^n)}{P_{\overline{X}^n}(X_\theta^n)} \le R - \frac{\gamma}{\sqrt{n}}\right\}.
\end{IEEEeqnarray}
From the definition of the $\nu$-typical set and (\ref{eq:lower2}), we also have
\begin{IEEEeqnarray}{rCl}  \label{eq:4-0-1}
\lefteqn{\liminf_{n \to \infty} \Pr\left\{ \frac{1}{n}\log \frac{P_{X_\theta^n}(X_{\theta}^n)}{P_{\overline{X}^n}(X_{\theta}^n)} \le R - \frac{\gamma}{\sqrt{n}} \right\}} \nonumber \\
& \ge & \liminf_{n \to \infty} \Pr\left\{ \frac{1}{n}\log \frac{P_{X_\theta^n}(X_{\theta}^n)}{P_{\overline{X}^n}(X_{\theta}^n)} \le R - \frac{\gamma}{\sqrt{n}}, \ X_\theta^n \in   {\cal T}_{\theta,\nu}^n\right\} \nonumber \\
& \ge & \liminf_{n \to \infty} \Pr\left\{   \frac{1}{n}\log \frac{P_{X_\theta^n}(X_{\theta}^n)}{\overline{P}^n_{\sigma(P_\theta)}(X_{\theta}^n)} \le R - \frac{\gamma}{\sqrt{n}}- \frac{1}{n} \log \frac{1}{c_{\tau_{m}}(P_{\theta})}  - \tau + \delta_\theta(\nu) \right\} \nonumber \\
&&- \limsup_{n \to \infty} \Pr\left\{ X_\theta^n \notin   {\cal T}_{\theta,\nu}^n\right\}\nonumber \\
& = & \liminf_{n \to \infty} \Pr\left\{   \frac{1}{n}\log \frac{P_{X_\theta^n}(X_{\theta}^n)}{\overline{P}^n_{\sigma(P_\theta)}(X_{\theta}^n)} \le R - \frac{\gamma}{\sqrt{n}}- \frac{1}{n} \log \frac{1}{c_{\tau_{m}}(P_{\theta})}  - \tau + \delta_\theta(\nu) \right\}
\end{IEEEeqnarray}
for any $\theta \in \Theta$.

We also partition the parameter space $\Theta$ into two sets.
\begin{IEEEeqnarray*}{rCl}
{\Theta}_1' &:=& \left\{ \theta \in \Theta  \left| D(P_\theta||\overline{P}_{\sigma(P_\theta)})  <  R  \right. \right\}, \\
{\Theta}_2' &:=& \left\{ \theta \in \Theta  \left| D(P_\theta||\overline{P}_{\sigma(P_\theta)})  \ge  R  \right. \right\}.
\end{IEEEeqnarray*}

Then, for $\theta \in \Theta_1'$, if we set $\nu>0$ and $\tau>0$ sufficiently small, then there exists a constant $\eta >0$ satisfying 
\begin{IEEEeqnarray}{rCl}
 R - \frac{\gamma}{\sqrt{n}}- \frac{1}{n} \log \frac{1}{c_{\tau_{m}}(P_{\theta})}  - \tau + \delta_\theta(\nu) > D(P_\theta||\overline{P}_{\sigma(P_\theta)}) + \eta \quad (\forall n > n_0).
\end{IEEEeqnarray}

Thus, again by invoking the weak law of large numbers, 
we have for $\forall \theta \in \Theta_1'$
\begin{IEEEeqnarray}{rCl}
\lefteqn{\liminf_{n \to \infty} \Pr\left\{ \frac{1}{n}\log \frac{P_{X_\theta^n}(X_\theta^n)}{\overline{P}^n_{\sigma(P_\theta)}(X^n_\theta)} \le R - \frac{\gamma}{\sqrt{n}}- \frac{1}{n} \log \frac{1}{c_{\tau_{m}}(P_{\theta})}  - \tau +\delta_\theta(\nu) \right\}} \nonumber \\
& \ge & \liminf_{n \to \infty} \Pr\left\{ \frac{1}{n}\log \frac{P_{X_\theta^n}(X_\theta^n)}{\overline{P}^n_{\sigma(P_\theta)}(X^n_\theta)} \le D(P_\theta||\overline{P}_{\sigma(P_\theta)}) + \eta \right\} \nonumber \\
& = & 1.
\end{IEEEeqnarray}

Summarizing up, we obtain
\begin{IEEEeqnarray}{rCl}
\lefteqn{\limsup_{n \to \infty} \Pr\left\{ \frac{1}{n}\log \frac{P_{X^n}(X^n)}{P_{\overline{X}^n}(X^n)} \le R \right\} } \nonumber \\
& \ge & \int_{\Theta} dw(\theta) \liminf_{n \to \infty} \Pr\left\{ \frac{1}{n}\log \frac{P_{X_\theta^n}(X_\theta^n)}{\overline{P}^n_{\sigma(P_\theta)}(X^n_\theta)} \le R - \frac{\gamma}{\sqrt{n}}- \frac{1}{n} \log \frac{1}{c_{\tau_{m}}(P_{\theta})}  - \tau + \delta_\theta(\nu) \right\} \nonumber \\
& \ge & \int_{\Theta_1'} dw(\theta) \liminf_{n \to \infty} \Pr\left\{ \frac{1}{n}\log \frac{P_{X_\theta^n}(X_\theta^n)}{\overline{P}^n_{\sigma(P_\theta)}(X^n_\theta)} \le D(P_\theta||\overline{P}_{\sigma(P_\theta)}) + \eta \right\}
 \nonumber \\
 & = & \int_{\Theta_1'} dw(\theta) \nonumber \\
 & = & \int_{\{\theta | D(P_\theta||\overline{P}_{\sigma(P_\theta)}) < R \} } dw(\theta).
\end{IEEEeqnarray}
This completes the proof of (\ref{eq:4-0-0-2}). 
\end{IEEEproof}
To illustrate the siginificance of Theorem \ref{theo:mixed2_1st}, let us now consider the special case with $\varepsilon=0$ and countably infinite parameter spaces as 
\begin{IEEEeqnarray*}{rCl}
\Theta & = & \left\{ 1,2,\cdots\right\}, \\
\Sigma &= &\left\{ 1,2,\cdots\right\}.
\end{IEEEeqnarray*}
In this case, we can write the \textit{null} and \textit{alternative} hypotheses $P_\theta$, $\overline{P}_\sigma$ as $P_i$, $\overline{P}_j$ with positive probability weights $w_i >0$, $v_j >0$ $(i,j=1,2,\cdots)$.
Then, by virtue of Theorem \ref{theo:mixed2_1st}, we have the following simplified result:
\begin{corollary} \label{coro:4-1}
For $\varepsilon=0$,
\begin{IEEEeqnarray}{rCl}  \label{eq:4-2-1}
B_0({\bf X}||\overline{\bf X}) = \inf_{i,j=1,2,\cdots} D(P_i||\overline{P}_j).
\end{IEEEeqnarray}
\end{corollary}
\begin{IEEEproof}

The formula (\ref{eq:theo:mixed2_1st}) can be written in this case as
\begin{IEEEeqnarray}{rCl} \label{eq:4-2-1-1}
B_0({\bf X}||\overline{\bf X}) = \sup \left\{ R \left| \Sigma_{\{ i|D(P_i||\overline{P}_{j(i)}) < R \}} w(i) = 0 \right.\right\},
\end{IEEEeqnarray}
where $j(i)$ is uniquely specified by
\begin{IEEEeqnarray}{rCl}
D(P_i||\overline{P}_{j(i)}) = \inf_{j=1,2,\cdots} D(P_i ||\overline{P}_j) \quad (i=1,2,\cdots),
\end{IEEEeqnarray}
because of the assumed closedness of $\Sigma$.
Let 
\begin{IEEEeqnarray*}{rCl}
R_1 < \sup\left\{ R | \Sigma_{\{i | D(P_i||\overline{P}_{j(i)}) <R \}} w(i) =0 \right\},
\end{IEEEeqnarray*}
then this means that
\begin{IEEEeqnarray}{rCl} \label{eq:4-2-2}
R_1 \le \inf_{i=1,2,\cdots} D(P_i||\overline{P}_{j(i)}).
\end{IEEEeqnarray}
Contrarily, let 
\begin{IEEEeqnarray*}{rCl}
R_2 > \sup\left\{ R | \Sigma_{\{i | D(P_i||\overline{P}_{j(i)}) <R \}} w(i) =0 \right\},
\end{IEEEeqnarray*}
then this means that 
\begin{IEEEeqnarray}{rCl} \label{eq:4-2-3}
R_2 \ge \inf_{i=1,2,\cdots} D(P_i||\overline{P}_{j(i)}).
\end{IEEEeqnarray}
As a consequence, (\ref{eq:4-2-1}) follows from (\ref{eq:4-2-1-1}), (\ref{eq:4-2-2}) and (\ref{eq:4-2-3}).
\end{IEEEproof}
\begin{remark}
One may wonder if it might be possible to deal with the second-order $\varepsilon$-optimum problem too using the arguments as developed in the above for the first-order $\varepsilon$-optimum problem with mixed memoryless sources ${\bf X}$ and $\overline{\bf X}$.
To do so, however, it seems that we need some novel techniques, which remains to be studied. 
\IEEEQED
\end{remark}
\section{hypothesis testing with mixed general sources}
We have so far considered the $\varepsilon$-hypothesis testing for mixed \textit{memoryless} sources.
In this section, we consider more general settings such as hypothesis testings with mixed \textit{general} sources. 

To do so, we consider the case where both of \textit{null} hypothesis ${\bf X}$ and \textit{alternative} hypothesis $\overline{\bf X}$ are \textit{finite} mixtures of \textit{general} sources as follows:
\begin{itemize}
\item The null hypothesis is a mixed \textit{general} source ${\bf X} = \{X^n \}_{n=1}^\infty$ consisting of $K$ \textit{general} (not necessarily memoryless) sources ${\bf X}_i = \{X^n_i \}_{n=1}^\infty\ (i=1,2,\cdots,K)$, that is, $\forall {\bf x} \in  {\cal X}^n$,
\begin{equation} \label{eq:mix5-1}
P_{X^n}({\bf x}) = \sum_{i=1}^K \alpha_iP_{X^n_i}({\bf x}),
\end{equation}
where $\alpha_i >0 \ (i=1,2,\cdots,K)$ and
$
\sum_{i=1}^K \alpha_i = 1.
$

\item The alternative hypothesis is another mixed \textit{general} source $\overline{\bf X}= \left\{\overline{X}^n \right\}_{n=1}^\infty$ consisting of $L$ \textit{general} (not necessarily memoryless) sources $\overline{{\bf X}}_j=\{\overline{X}_{j}^n\}_{n=1}^\infty \ (j=1,\cdots,L)$, that is, $\forall {\bf x} \in  {\cal X}^n$,
\begin{equation} \label{eq:mix5-2}
P_{\overline{X}^n}({\bf x}) = \sum_{j=1}^L \beta_jP_{\overline{X}^n_j}({\bf x}),
\end{equation}
where $\beta_j >0 \ (j=1,2,\cdots,L)$ and 
$
\sum_{j=1}^L \beta_j = 1.
$
\end{itemize}
In this general setting, it is hard to derive a compact formula for the first-order $\varepsilon$-optimum exponent (for $0 \le \varepsilon <1$). 
Instead, we can obtain the following theorem in the special case of $\varepsilon=0$. 
\begin{theorem}  \label{theo:5-1}
\begin{IEEEeqnarray}{rCl}  \label{eq:5-2-1-1}
B_0({\bf X}||\overline{\bf X}) = \min_{1\le i \le K, 1 \le j \le L} B_0({\bf X}_i||\overline{\bf X}_j).
\end{IEEEeqnarray}
In particular, if ${\bf X}_i$ and $\overline{\bf X}_j$ are all stationary memoryless sources specified by $X_i \ (i=1,2,\cdots,K)$ and $\overline{X}_j \ (j=1,2,\cdots,L)$, respectively,
then
\[
B_0({\bf X}||\overline{\bf X}) = \min_{1\le i \le K, 1 \le j \le L} D(P_{ X_i}||P_{\overline{ X}_j}),
\]
 which is a special case of Corollary \ref{coro:4-1}.
\end{theorem}
\begin{IEEEproof}

The proof proceeds in parallel with the argument in \cite[Remark 4.4.3]{Han}.
To be self-contained, we fully describe it in Appendix B. 
\end{IEEEproof}
We can consider the following exponentially $r$-optimum exponent in the hypothesis testing with the two mixed \textit{general} sources ${\bf X}$ and $\overline{\bf X}$ as above.
\begin{definition} \label{def:r}
Let $r >0$ be any fixed constant. Rate $R$ is said to be exponentially $r$-achievable if there exists an acceptance region ${\cal A}_n$ such that
\begin{IEEEeqnarray*}{rCl}
\liminf_{n \to \infty} \frac{1}{n}\log \frac{1}{\mu_n} & \ge & r, \\
\liminf_{n \to \infty} \frac{1}{n}\log \frac{1}{\lambda_n} & \ge & R.
\end{IEEEeqnarray*}
\end{definition}
\begin{definition}[First-order exponentially $r$-optimum exponent]
\begin{equation}
B_e(r|{\bf X}||\overline{\bf X}):= \sup\{ R | R \mbox{ is exponentially $r$-achievable} \}.
\end{equation}
\end{definition}
Then, it is not difficult to verify that an analogous result to Theorem \ref{theo:5-1} holds (which is a generalization of \cite[Remark 4.4.3]{Han}):
\begin{theorem} \label{theo:r1}
\begin{IEEEeqnarray*}{rCl} 
B_e(r|{\bf X}||\overline{\bf X}) = \min_{1\le i \le K, 1 \le j \le L} B_e(r|{\bf X}_i||\overline{\bf X}_j).
\end{IEEEeqnarray*}
In particular, if the \textit{null} and \textit{alternative} hypotheses consist of stationary memoryless sources $X_i \ (i=1,2,\cdots,K)$ and $\overline{X}_j \ (j=1,2,\cdots,L)$, respectively, then 
\begin{IEEEeqnarray}{rCl} \label{eq:5-2-1}
B_e(r|{\bf X}||\overline{\bf X}) = \min_{1\le i \le K, 1 \le j \le L}\inf_{P_{\tilde{X}}:D(P_{\tilde{X}}||P_{X_i}) < r} D(P_{\tilde{X}}||P_{\overline{X}_j}),
\end{IEEEeqnarray}
by virtue of Hoeffding's theorem. 
\IEEEQED
\end{theorem}
%
%
%
%
%
%
 \label{sec:compound}
\section{Hypothesis testing with compound sources} 
In this section, we consider the \textit{compound} hypothesis testing problem with finite \textit{null} hypotheses ${\bf X}_i = \{X_i^n \}_{n=1}^\infty \ (i=1,2,\cdots,K)$ and finite \textit{alternative} hypotheses $\overline{\bf X}_j = \{ \overline{X}_j^n \}_{n=1}^\infty \ (j=1,2,\cdots, L)$, where ${\bf X}_i$ and $\overline{\bf X}_j$ are \textit{general} sources. 

The {compound} hypothesis testing is the problem in which a pair  of \textit{general} sources $({\bf X}_i,\overline{\bf X}_j)$ occurs as a pair (\textit{null} hypothesis, \textit{alternative} hypothesis), and the tester does not know which pair $({\bf X}_i,\overline{\bf X}_j)$ is actually working.
This means that the acceptance region ${\cal A}_n$ cannot depend on $i$ and $j$. 
The type I error of the {compound} hypothesis testing is given by
\begin{equation}
\mu_n^{(i)} := \Pr\left\{ X^n_i \notin {\cal A}_n \right\},
\end{equation}
for each \textit{general} null hypothesis ${\bf X}_i$.
The type II error is also given by
\begin{equation}
\lambda_n^{(j)} := \Pr\left\{ \overline{X}^n_j \in {\cal A}_n \right\},
\end{equation}
for each \textit{general} alternative hypothesis $\overline{\bf X}_j$.
Then, the following achievability is of our interest.
\begin{definition}
Rate $R$ is said to be $0$-achievable for the compound hypothesis testing, if there exists an acceptance region ${\cal A}_n$ such that
\begin{equation*} 
\lim_{n \to \infty} \mu_n^{(i)} = 0 \   \mbox{ and } \ \liminf_{n \to \infty} \frac{1}{n}\log \frac{1}{\lambda_n^{(j)}} \ge R,
\end{equation*}
for all $i=1,2,\cdots,K$ and $j=1,2,\cdots,L$.
\end{definition}
\begin{definition}[First-order optimum exponent]
\begin{equation}
B(\{{\bf X}_i\}_{i=1}^K||\{\overline{\bf X}_j\}_{j=1}^L) := \sup\{ R | R \mbox{ is $0$-achievable} \}.
\end{equation}
\end{definition}
The following theorem reveals the relationship between the hypothesis testing with mixed \textit{general} sources as defined in (\ref{eq:mix5-1}) and (\ref{eq:mix5-2}), and the compound hypothesis testing with the \textit{general} sources.
\begin{theorem} \label{theo:compound}
Assuming that $\alpha_i > 0$ and $\beta_j > 0$ hold for all $ i=1,2,\cdots,K$ and $ j=1,2,\cdots,L$, it holds that
\begin{IEEEeqnarray}{rCl}
B(\{{\bf X}_i\}_{i=1}^K||\{\overline{\bf X}_j\}_{j=1}^L) = B(\{\alpha_i,{\bf X}_i\}_{i=1}^K||\{ \beta_j,\overline{\bf X}_j \}_{j=1}^L),
\end{IEEEeqnarray}
where with sources (\ref{eq:mix5-1}) and (\ref{eq:mix5-2}) we use here the notation
\begin{IEEEeqnarray}{rCl} \label{eq:5-2-3}
B(\{\alpha_i,{\bf X}_i\}_{i=1}^K||\{ \beta_j,\overline{\bf X}_j \}_{j=1}^L) 
\end{IEEEeqnarray}
to denote $B_0({\bf X}||\overline{\bf X})$ to make explicit the dependence on $\alpha_i$, $\beta_j$.
\end{theorem}
\begin{IEEEproof}

It suffices to show two inequalities:
\begin{IEEEeqnarray}{rCl} \label{eq:6-1}
B(\{{\bf X}_i\}_{i=1}^K||\{\overline{\bf X}_j\}_{j=1}^L) & \le B(\{\alpha_i,{\bf X}_i\}_{i=1}^K||\{ \beta_j,\overline{\bf X}_j \}_{j=1}^L), \\ \label{eq:6-1-2}
B(\{{\bf X}_i\}_{i=1}^K||\{\overline{\bf X}_j\}_{j=1}^L) & \ge B(\{\alpha_i,{\bf X}_i\}_{i=1}^K||\{ \beta_j,\overline{\bf X}_j \}_{j=1}^L).
\end{IEEEeqnarray}

\textit{Proof of (\ref{eq:6-1})}:
Suppose that $R$ is $0$-achievable for the compound hypothesis testing, that is, there exists an acceptance region ${\cal A}_n$ such that
\begin{IEEEeqnarray}{rCl} \label{eq:6-2}
\lim_{n \to \infty} \mu_n^{(i)} & = & 0 \quad (i=1,2,\cdots,K), \\ \label{eq:6-3}
\liminf_{n \to \infty} \frac{1}{n}\log \frac{1}{\lambda_n^{(j)}} &\ge & R \quad (j=1,2,\cdots,L).
\end{IEEEeqnarray}
Then, the type I error probability $\mu_n$ for the hypothesis testing with mixed \textit{general} sources is evaluated as follows.
By the definition of $\mu_n$ and (\ref{eq:mix5-1}), we have
\begin{IEEEeqnarray*}{rCl}
\mu_n  & = & \Pr \left\{X^n \notin {\cal A}_n \right\}\\ 
& = & \sum_{i=1}^K \alpha_i \Pr\left\{ X^n_i \notin {\cal A}_n \right\} \\
& = & \sum_{i=1}^K \alpha_i\mu_n^{(i)},
\end{IEEEeqnarray*}
from which, together with (\ref{eq:6-2}), we obtain
\begin{equation} \label{eq:6-4}
\lim_{n \to \infty} \mu_n = 0.
\end{equation}
Similarly, we have
\begin{IEEEeqnarray}{rCl} \label{eq:6-5}
\lambda_n & = &  \Pr\left\{ \overline{X}^n \in {\cal A}_n\right\} \nonumber \\
& = &  \sum_{j=1}^L \beta_j\Pr\left\{ \overline{X}_j^n \in {\cal A}_n\right\} \nonumber \\
& = &  \sum_{j=1}^L \beta_j \lambda_n^{(j)}.
\end{IEEEeqnarray}
On the other hand, (\ref{eq:6-3}) implies
\begin{IEEEeqnarray*}{rCl}
{\lambda_n^{(j)}} &\le & e^{n(R-\gamma)} \quad (n \ge n_0),
\end{IEEEeqnarray*}
holds for any $\gamma >0$ and all $j=1,2,\cdots,L$.
Substituting this inequality into (\ref{eq:6-5}) yields
\begin{IEEEeqnarray}{rCl} \label{eq:6-6}
\liminf_{n \to \infty} \frac{1}{n}\log \frac{1}{\lambda_n} & \ge R - \gamma.
\end{IEEEeqnarray}
Since $\gamma >0$ is arbitrary, from (\ref{eq:6-4}) and (\ref{eq:6-6}) we conclude that (\ref{eq:6-1}) holds.

\textit{Proof of (\ref{eq:6-1-2})}:

Suppose that $R$ is $0$-achievable for the mixed hypothesis testing, that is, there exists an acceptance region ${\cal A}_n$ such that
\begin{IEEEeqnarray}{rCl} \label{eq:6-7}
\lim_{n \to \infty} \mu_n & = & 0, \\ \label{eq:6-8}
\liminf_{n \to \infty} \frac{1}{n}\log \frac{1}{\lambda_n} &\ge & R .
\end{IEEEeqnarray}
We fix such an ${\cal A}_n$ and set
\begin{IEEEeqnarray*}{rCl}
\mu_n^{(i)} & = \Pr\left\{ X_i^n \notin {\cal A}_n\right\},  \\ 
\lambda_n^{(j)}& = \Pr\left\{ \overline{X}_j^n \in {\cal A}_n\right\}.
\end{IEEEeqnarray*}
Then, from (\ref{eq:mix5-1}) we have
\begin{IEEEeqnarray*}{rCl}
\mu_n  & = & \sum_{i=1}^K \alpha_i \Pr\left\{ X_i^n \notin {\cal A}_n\right\} \\
& = & \sum_{i=1}^K \alpha_i \mu_n^{(i)},
\end{IEEEeqnarray*}
from which, it follows that
\[
\mu_n^{(i)} \le \frac{\mu_n}{\alpha_i}
\]
for all $i = 1,2,\cdots,K$.
From this inequality and (\ref{eq:6-7}), we obtain
\begin{IEEEeqnarray}{rCl} \label{eq:6-9}
\lim_{n \to \infty} \mu_n^{(i)} & = & 0,
\end{IEEEeqnarray}
for all $i =1,2,\cdots,K$.
Similarly, 
\begin{IEEEeqnarray*}{rCl}
\lambda_n & = & \sum_{j=1}^L \beta_j \Pr \left\{ \overline{X}^n_j \in {\cal A}_n \right\} \\
& = & \sum_{j=1}^L \beta_j \lambda_n^{(j)},
\end{IEEEeqnarray*}
so that we have for $j=1,2,\cdots,L$,
\[
\lambda_n^{(j)} \le \frac{\lambda_n}{\beta_j},
\]
which means that
\begin{IEEEeqnarray*}{rCl}
\frac{1}{n}\log \frac{1}{\lambda_n^{(j)}} & \ge & \frac{1}{n}\log \frac{\beta_j}{\lambda_n} \nonumber \\
& = & \frac{1}{n}\log \frac{1}{\lambda_n} - \frac{1}{n}\log \frac{1}{\beta_j}.
\end{IEEEeqnarray*}
Noting that $\beta_j \ (j=1,2,\cdots,L)$ are constants, from (\ref{eq:6-8}) we obtain
\begin{IEEEeqnarray}{rCl} \label{eq:6-10}
\liminf_{n \to \infty} \frac{1}{n}\log \frac{1}{\lambda_n^{(j)}} &\ge & R,
\end{IEEEeqnarray}
for all $j=1,2,\cdots,L$.
From (\ref{eq:6-9}) and (\ref{eq:6-10}), we conclude that (\ref{eq:6-1-2}) holds.
\end{IEEEproof}

From Theorems \ref{theo:5-1} and \ref{theo:compound}, we immediately obtain the first-order $0$-optimum exponent for the \textit{compound} hypothesis testing as:
\begin{corollary} \label{coro:compound}
Assuming that $\alpha_i > 0$ and $\beta_j > 0$ hold for all $i=1,2,\cdots,K$ and $j=1,2,\cdots,L$, we have
\begin{IEEEeqnarray}{rCl}  \label{eq:5-2-4}
B(\{{\bf X}_i\}_{i=1}^K||\{\overline{\bf X}_j\}_{j=1}^L) = \min_{1\le i \le K, 1 \le j \le L} B_0({\bf X}_i||\overline{\bf X}_j).
\end{IEEEeqnarray}
In particular, if ${\bf X}_i$ and $\overline{\bf X}_j$ are all stationary memoryless sources specified by $X_i$ and $\overline{X}_j$, respectively, (\ref{eq:5-2-4}) reduces to 
\[
B(\{{\bf X}_i\}_{i=1}^K||\{\overline{\bf X}_j\}_{j=1}^L) = \min_{1\le i \le K, 1 \le j \le L} D(P_{ X_i}||P_{\overline{ X}_j}).
\]
\IEEEQED
\end{corollary}
\begin{remark}
Similarly to Definition \ref{def:r}, we can define the exponentially $r$-optimum exponent also for the {compound} hypothesis testing problem as follows.
\begin{definition}
Let $r >0$ be any fixed constant.
Rate $R$ is said to be exponentially $r$-achievable for the compound hypothesis testing, if there exists an acceptance region ${\cal A}_n$ such that
\begin{IEEEeqnarray*}{rCl} 
\liminf_{n \to \infty} \frac{1}{n}\log \frac{1}{\mu_n^{(i)}} & \ge & r, \\
\liminf_{n \to \infty} \frac{1}{n}\log \frac{1}{\lambda_n^{(j)}} & \ge & R,
\end{IEEEeqnarray*}
for all $i=1,2,\cdots,K$ and $j=1,2,\cdots,L$.
\end{definition}
\begin{definition}[First-order exponentially $r$-optimum exponent]
\begin{equation}
B_e(r|\{{\bf X}_i\}_{i=1}^K||\{\overline{\bf X}_j\}_{j=1}^L) := \sup\{ R | R \mbox{ is exponentially $r$-achievable} \}.
\end{equation}
\end{definition}
Then, using a similar argument to the proof of Theorem \ref{theo:compound}, the following theorem can be shown:
\begin{theorem} \label{theo:compound2}
Let $\alpha_i > 0$ and $\beta_j > 0$ hold for all $ i=1,2,\cdots,K$ and $  j=1,2,\cdots,L$, then it holds that
\begin{IEEEeqnarray*}{rCl}
B_e(r|\{{\bf X}_i\}_{i=1}^K||\{\overline{\bf X}_j\}_{j=1}^L) = B_e(r|\{\alpha_i,{\bf X}_i\}_{i=1}^K||\{ \beta_j,\overline{\bf X}_j \}_{j=1}^L),
\end{IEEEeqnarray*}
where  we use the notation
\begin{IEEEeqnarray*}{rCl} 
B_e(r|\{\alpha_i,{\bf X}_i\}_{i=1}^K||\{ \beta_j,\overline{\bf X}_j \}_{j=1}^L)
\end{IEEEeqnarray*}
to denote $B_e(r|{\bf X}||\overline{\bf X})$. 
\end{theorem}
Combining Theorems \ref{theo:r1} and \ref{theo:compound2}, we immediately obtain the following corollary.
\begin{corollary}
Let $\alpha_i > 0$ and $\beta_j > 0$ hold for all $ i=1,2,\cdots,K$ and $  j=1,2,\cdots,L$, then it holds that
\begin{IEEEeqnarray*}{rCl}
B_e(r|\{{\bf X}_i\}_{i=1}^K||\{\overline{\bf X}_j\}_{j=1}^L) = \min_{1\le i \le K, 1 \le j \le L} B_e(r|{\bf X}_i||\overline{\bf X}_j).
\end{IEEEeqnarray*}
In particular, if the null and alternative hypotheses consist of stationary memoryless sources specified by $X_i \ (i = 1,2,\cdots, K)$ and $\overline{X}_j\ (j = 1,2,\cdots, L)$, respectively,  as in Theorem \ref{theo:r1}, then 
\begin{IEEEeqnarray*}{rCl}
B_e(r|\{{\bf X}_i\}_{i=1}^K||\{\overline{\bf X}_j\}_{j=1}^L) =  \min_{1\le i \le K, 1 \le j \le L} \inf_{P_{\tilde{X}}:D(P_{\tilde{X}}||P_{X_i}) < r} D(P_{\tilde{X}}||P_{\overline{X}_j}),
\end{IEEEeqnarray*}
which corresponds to (\ref{eq:5-2-1}).
\end{corollary}
\IEEEQED
\end{remark}

\section{concluding remarks}
So far, we have considered the \textit{first-} and \textit{second}-order $\varepsilon$-optimum exponents in the hypothesis testing problem.
First, we have studied the second-order $\varepsilon$-optimum problem with mixed memoryless sources.
As we have shown in the analysis of the second-order $\varepsilon$-optimum exponent, we use, as a key property, the asymptotic normality of \textit{divergence density rate} for each of the component sources. 
We also observe that the \textit{canonical representation}, first introduced in \cite{NH2014}, is still efficient to express the second-order $\varepsilon$-optimum exponent for mixed memoryless sources in the hypothesis testing problem.

The \textit{first-}order $\varepsilon$-optimum exponent in the case with mixed memoryless \textit{null} and \textit{alternative} hypotheses has also been established.
One may wonder whether we can apply the same approach in the derivation of the \textit{second-}order $\varepsilon$-optimum exponent in this setting.
One of our key techniques to derive the \textit{first-}order $\varepsilon$-optimum exponent
is an expansion $P_{\bf x}$ around $P_\theta$.
More careful evaluation of this expansion would be needed to compute the second-order $\varepsilon$-optimum exponent.
This is a future work.

The relationship between the \textit{first}-order $0$-optimum (resp. exponentially $r$-optimum) exponent in the hypothesis testing with mixed \textit{general} sources and the $0$-optimum (resp. exponentially $r$-optimum) exponent in the \textit{compound} hypothesis testing has also been demonstrated.
%
%
%
%
%
%
%

%

\appendices
\section{Proof of Lemmas \ref{upper} and \ref{lower}}
\renewcommand{\theequation}{A.\arabic{equation}}
\setcounter{equation}{0}
\textit{Proof of Lemma \ref{upper}:}

Since 
\begin{IEEEeqnarray*}{rCl}
1 
& \ge & \sum_{{\bf x} \in {\cal A}_n} P_{X^n}({\bf x}) \\
& \ge & \sum_{{\bf x} \in {\cal A}_n} P_{\overline{X}^n}({\bf x}) e^{nt} \\
& = & \Pr \left\{ \overline{X}^n \in {\cal A}_n \right\} e^{nt}
\end{IEEEeqnarray*}
holds, we obtain
\begin{IEEEeqnarray*}{rCl}
\Pr \left\{ \overline{X}^n \in {\cal A}_n \right\} \le e^{nt},
\end{IEEEeqnarray*}
which completes the proof of the lemma.

\textit{Proof of Lemma \ref{lower}:}

Define 
\begin{equation*}
S_n := \left\{ {\bf x} \in {\cal X}^n \left| \frac{1}{n} \log \frac{P_{X^n}({\bf x})}{P_{\overline{X}^n}({\bf x})} \le t \right. \right\}.
\end{equation*}
Then, it follows that
\begin{IEEEeqnarray}{rCl}  \label{eq:app1}
\Pr \left\{ \frac{1}{n} \log \frac{P_{{X}^n}(X^n)}{P_{\overline{X}^n}(X^n)} \le t \right\} & = & \Pr \left\{ X^n \in S_n \right\} \nonumber \\
& = & \Pr \left\{ X^n \in S_n \cup {\cal A}_n^c  \right\} +  \Pr \left\{ X^n \in S_n \cup {\cal A}_n \right\} \nonumber \\
& \le & \Pr \left\{ X^n \in {\cal A}_n^c  \right\} + \Pr \left\{ X^n \in S_n \cup {\cal A}_n \right\}
\end{IEEEeqnarray}
for any ${\cal A}_n \subset {\cal X}^n$, where $D^c$ denotes the complement of $D$.
The second term on the right-hand side of (\ref{eq:app1}) is upper bounded as
\begin{IEEEeqnarray}{rCl}  \label{eq:app2}
\Pr \left\{ X^n \in S_n \cup {\cal A}_n \right\} & = & \sum_{{\bf x} \in S_n \cup{\cal A}_n} P_{X^n}({\bf x}) \nonumber \\
& \le & \sum_{{\bf x} \in S_n \cup {\cal A}_n} P_{\overline{X}^n}({\bf x}) e^{nt} \nonumber \\
& \le & \sum_{{\bf x} \in  {\cal A}_n} P_{\overline{X}^n}({\bf x}) e^{nt} \nonumber \\
& = & e^{nt} \Pr\left\{ \overline{X}^n \in {\cal A}_n \right\}.
\end{IEEEeqnarray}
Substituting (\ref{eq:app2}) into (\ref{eq:app1}), we have the lemma.
%
%
%
%
%
%
%
\section{Proof of Theorem \ref{theo:5-1}}
\renewcommand{\theequation}{B.\arabic{equation}}
\setcounter{equation}{0}
First, we prove the inequality:
\begin{IEEEeqnarray}{rCl}  \label{eq:5-0}
B_0({\bf X}||\overline{\bf X}) \ge \min_{ 1\le i \le K, 1 \le j \le L} B_0({\bf X}_i||\overline{\bf X}_j).
\end{IEEEeqnarray}
To do so, we arbitrarily fix $R_{ij}$ for $1 \le \forall i \le K,1 \le \forall j \le L$ so that
\begin{equation} \label{eq:5-1-2}
R_{ij} < B_0({\bf X}_i||\overline{\bf X}_j).
\end{equation}
Then, by the definition of $B_0({\bf X}_i||\overline{\bf X}_j)$, there exists an acceptance region ${\cal A}_n^{(i,j)}$ satisfying
\begin{IEEEeqnarray}{rCl} \label{eq:5-2}
\lim_{n \to \infty} \mu_n^{(i,j)} & = 0, \\ \label{eq:5-2-2}
\liminf_{n \to \infty} \frac{1}{n}\log \frac{1}{\lambda_n^{(i,j)}} & \ge R_{ij}, 
\end{IEEEeqnarray}
where $\mu_n^{(i,j)}$ and $\lambda_n^{(i,j)}$ are defined respectively as 
\begin{IEEEeqnarray*}{rCl}
\mu_n^{(i,j)} := \Pr\left\{X^n \notin {\cal A}_n^{(i,j)}\right\}, \quad \lambda_n^{(i,j)} := \Pr\left\{\overline{X}^n \in {\cal A}_n^{(i,j)}\right\}.
\end{IEEEeqnarray*}
By using these regions, we define the acceptance region ${\cal A}_n$ as 
\begin{equation}
{\cal A}_n := \bigcup_{i=1}^K \left(\bigcap_{j=1}^L{\cal A}_n^{(i,j)}\right).
\end{equation}
Then, we have
\begin{IEEEeqnarray*}{rCl}
\mu_n = \Pr \left\{X^n \notin {\cal A}_n \right\} 
& = & \sum_{i=1}^K \alpha_i  \Pr\left\{ X^n_i \notin \left(\bigcup_{i=1}^{K} \bigcap_{j=1}^L {\cal A}_n^{(i,j)} \right) \right\}\\
& \le & \sum_{i=1}^K \alpha_i \Pr\left\{ X^n_i \notin \left(\bigcap_{j=1}^L {\cal A}_n^{(i,j)} \right) \right\} \\
& \le & \sum_{i=1}^K  \sum_{j=1}^L \alpha_i \Pr\left\{ X^n_i \notin \left({\cal A}_n^{(i,j)} \right) \right\} \\
& = & \sum_{i=1}^K \sum_{j=1}^L \alpha_i\mu_n^{(i,j)},
\end{IEEEeqnarray*}
from which, together with (\ref{eq:5-2}), we obtain
\begin{equation} \label{eq:5-3}
\lim_{n \to \infty} \mu_n = 0.
\end{equation}
Similarly, we have
\begin{IEEEeqnarray*}{rCl}
\lambda_n = \Pr\left\{ \overline{X}^n \in {\cal A}_n \right\} & \le &  \sum_{j=1}^L \sum_{i=1}^K \beta_j \lambda_n^{(i,j)} ,
\end{IEEEeqnarray*}
from which, together with (\ref{eq:5-2-2}), we obtain
\begin{IEEEeqnarray*}{rCl}
\liminf_{n \to \infty} \frac{1}{n}\log \frac{1}{\lambda_n} & \ge  \min_{1\le i \le K, 1 \le j \le L} R_{ij}.
\end{IEEEeqnarray*}
Since $R_{ij}$ are arbitrary as far as (\ref{eq:5-1-2}) is satisfied, we have (\ref{eq:5-0}).

Next, we prove the inequality:
\begin{IEEEeqnarray}{rCl}  \label{eq:5-4}
B_0({\bf X}||\overline{\bf X}) \le \min_{1\le i \le K, 1 \le i \le L} B_0({\bf X}_i||\overline{\bf X}_j).
\end{IEEEeqnarray}
To do so, let $R$ be $0$-achievable, then there exists an acceptance region ${\cal A}_n$ satisfying
\begin{IEEEeqnarray}{rCl} \label{eq:5-5-1}
\lim_{n \to \infty} \mu_n & = 0, \\ \label{eq:5-5-2}
\liminf_{n \to \infty} \frac{1}{n}\log \frac{1}{\lambda_n} & \ge R. 
\end{IEEEeqnarray}
We fix such an ${\cal A}_n$ and consider the hypothesis testing with \textit{null} hypothesis ${\bf X}_i$ and \textit{alternative} hypothesis $\overline{\bf X}_j$ for arbitrarily fixed $i$ and $j$.
Then, probabilities of type I error and type II error are given by
\begin{IEEEeqnarray*}{rCl}
\mu_n^{(i,j)} & = & \Pr \left\{ X^n_i \notin {\cal A}_n \right\}, \\
\lambda_n^{(i,j)} & = & \Pr \left\{ \overline{X}_j^n \in {\cal A}_n \right\}.
\end{IEEEeqnarray*}
Since
\begin{IEEEeqnarray*}{rCl}
\mu_n & = & \sum_{i=1}^K \alpha_i \Pr \left\{ X^n_i \notin {\cal A}_n \right\} \\
& = &  \sum_{i=1}^K \alpha_i \mu_n^{(i,j)},
\end{IEEEeqnarray*}
we have 
\begin{IEEEeqnarray}{rCl}  \label{eq:5-6-0}
\mu_n^{(i,j)} \le \frac{\mu_n}{\alpha_i}.
\end{IEEEeqnarray}
From this inequality and (\ref{eq:5-5-1}) we obtain
\begin{IEEEeqnarray}{rCl} \label{eq:5-6-1}
\lim_{n \to \infty} \mu_n^{(i,j)} = 0. 
\end{IEEEeqnarray}

Similarly to the derivation of (\ref{eq:5-6-0}), we have
\begin{IEEEeqnarray*}{rCl} 
\lambda_n^{(i,j)} \le \frac{\lambda_n}{\beta_j}.
\end{IEEEeqnarray*}
Hence, from (\ref{eq:5-5-2}) we obtain
\begin{IEEEeqnarray}{rCl} \label{eq:5-6-2}
R & \le & \liminf_{n \to \infty} \frac{1}{n}\log \frac{1}{\lambda_n} \nonumber \\
& \le & \liminf_{n \to \infty} \frac{1}{n}\log \frac{1}{\lambda_n^{(i,j)}} + \limsup_{n \to \infty} \frac{1}{n}\log \frac{1}{\beta_j} \nonumber \\
& = & \liminf_{n \to \infty} \frac{1}{n}\log \frac{1}{\lambda_n^{(i,j)}}.
\end{IEEEeqnarray}
From (\ref{eq:5-6-1}) and (\ref{eq:5-6-2}), it follows that $R$ is $0$-achievable for the hypothesis testing with ${\bf X}_i$ against $\overline{\bf X}_j$. 
Noting that $i,j$ are arbitrary with $1 \le i \le K$ and $1\le j \le L$,
we obtain
\begin{IEEEeqnarray*}{rCl} 
R \le \min_{1\le i \le K, 1 \le i \le L} B_0({\bf X}_i||\overline{\bf X}_j).
\end{IEEEeqnarray*}
This means that  (\ref{eq:5-4}) holds, completing the proof of Theorem \ref{theo:5-1}.
\IEEEQED



\ifCLASSOPTIONcaptionsoff
  \newpage
\fi

\end{document}